\documentclass[pdftex]{article}
\usepackage[top=25truemm,bottom=20truemm,left=20truemm,right=20truemm]{geometry}
\usepackage{authblk}

\usepackage{abstract}
\usepackage{amsmath}
\usepackage{amssymb}
\usepackage{amsfonts}

\usepackage{here}
\usepackage{tikz}
\usepackage{cite}
\usepackage{bm}
\usepackage{setspace}
\usepackage{enumerate}
\usepackage{arydshln}
\usepackage{caption}
\usepackage{color}
\usepackage{appendix}
\usepackage{lineno}
\usepackage{ulem}

\title{\textbf{A particle-based method using the mesh-constrained discrete point approach for two-dimensional Stokes flows}}

\author[1]{Takeharu MATSUDA}
\author[1]{Kohsuke TSUKUI}
\author[1]{Satoshi II\footnote{Corresponding Author\\~~~~ Email: sii@tmu.ac.jp (S. Ii)}}

\affil[1]{Graduate School of Systems Design, Tokyo Metropolitan University, 1-1 Minami-Osawa, Hachioji, Tokyo 192-0397, Japan}

\begin{document}

\date{}
\maketitle

\begin{abstract}
  \noindent Meshless methods inherently do not require mesh topologies and are practically used for solving continuum equations. However, these methods generally tend to have a higher computational load than conventional mesh-based methods because calculation stencils for spatial discretization become large. In this study, a novel approach for the use of compact stencils in meshless methods is proposed, called the mesh-constrained discrete point (MCD) approach. The MCD approach introduces a  Cartesian mesh system  to the background of a domain.  And the approach rigorously constrains the distribution of discrete points (DPs) in each mesh by solving a dynamic problem with nonlinear constraints. This can avoid the heterogeneity of the DP distribution at the mesh-size level and impose compact stencils with a fixed degree of freedom for derivative evaluations. A fundamental formulation for arrangements of DPs and an application to unsteady Stokes flows are presented in this paper. Numerical tests were performed for the distribution of DPs and flow problems in co-axial and eccentric circular channels. The proposed MCD approach achieved a reasonable distribution of DPs independently of the spatial resolution with a few iterations in pre-processing. Additionally, solutions using the obtained DP distributions in Stokes flow problems were in good agreement with theoretical and reference solutions. The results also confirmed that the numerical accuracies of velocity and pressure achieved the expected convergence order, even when compact stencils were used.
  \vskip.5\baselineskip
\end{abstract}


\section{Introduction} \label{introduction}

Meshless (or meshfree) methods belong to a class of numerical schemes for solving partial differential equations (PDEs) in continuum mechanics.
The greatest advantage of these methods is that there is no requirement for cumbersome mesh generation.
The methods are mainly classified into two types of formulations: Galerkin and collocation types.
Generally, the Galerkin-type formulation is superior in terms of numerical accuracy and stability because of the weak form of PDEs.
Several outstanding Galerkin-type formulations exist (Nayroles et al., 1992; Belytschko et al., 1994; Liu et al., 1995; Melenk and Babuska, 1999; Chen et al., 2013), and many variations.
In the Galerkin-type formulation, background meshes/grids or auxiliary domain subdivision are introduced for the purpose of numerical integration.
By contrast, the collocation-type formulation is typically derived using a strong form of PDEs with the Dirac delta; thus, a mesh is not required for the core of the formulation, which results in a pure meshless method. 
There are also several outstanding collocation-type formulations (Onate et al., 1996; Zhang et al., 2001; Afshar and Lashckarbolok 2008).

Meshless methods that adopt the Lagrangian description of material motion are also called (Lagrangian) particle methods, in which discrete points (DPs) for unknowns are defined as particles and moved according to the material motion.
There are two pioneering methods in computational fluid dynamics: the smoothed particle hydrodynamics method (Monaghan, 1992) and moving particle semi-implicit/simulation (MPS) method (Koshizuka and Oka, 1996).
These methods powerfully solve complex problems, including multiphase flows, interfacial flows, and fluid-structure interaction problems with large deformations (Cummins and Rudman, 1999; Colagrossi and Landrini, 2003; Tanaka et al., 2018; Zhang et al., 2021; Li et al., 2022; Shimizu et al., 2022).
Regarding high-order discretization, using the idea of the moving least-squares (MLS) method (Lancaster and Salkauskas, 1981), high-precision particle methods that ensure an arbitrary order of spatial accuracy have been proposed and called the MLS reproducing kernel  method (Liu et al., 1997) and least-squares MPS (LSMPS) method (Tamai and Koshizuka, 2014). Both methods introduce a scaling parameter to avoid ill-posedness in the polynomial reconstruction; however, the LSMPS method achieves more suitable scaling and is stable when solving a linear system compared with conventional methods. 
Recently, the LSMPS method was extended to improve the pressure disturbances inherent to particle methods by formulating a spatial discretization scheme for the LSMPS method that considers Neumann boundary conditions (Matsunaga et al., 2020).

In meshless methods, a local configuration of DPs is required to evaluate spatial derivatives using any approximation.
In this regard, the local number densities of the DPs (or particles in particle methods) are not constant in space; hence, broad and numerous stencils for spatial discretization should be retained to avoid ill-posedness in derivative evaluations that increase the algebraic manipulation and non-zero components of coefficient matrices in linear systems of discretized governing equations.
This issue is the reason that meshless methods are inferior to mesh-based methods in terms of computational efficiency, for example, finite difference and finite element methods.
Additionally, the heterogeneous distribution of DPs causes another issue of the imbalance of the computational load in parallelization for large-scale simulations.
From a practical point of view, background meshes/grids (or buckets) that encompass a computational domain are introduced to evaluate the local configuration of DPs, where each DP is linked to any background mesh, and it is easy to access the local (or surrounding) DPs from any DP.
An efficient parallelization technique was proposed (Murotani et al., 2015) that uses these background meshes.
However, it is competitive to achieve a good balance between computational cost in each parallelization node and node-to-node communication, and it seems to be difficult to develop a universal technique.

As a different direction with the full particle methods, several hybrid methods using both the particles/DPs and meshes have been proposed (Brackbill and Ruppel, 1986; Liu et al., 2005; Zhang and Liu, 2009; Matsunaga et al., 2015).
Most of these hybrid methods are formulated to address coupled behaviors of multiple materials in multiphase flows or detailed flow motions in sub-cell resolution, where the particles are used to track the material motions and the meshes are used to solve the continuum equations.
Although the numerical accuracy and practicality have been shown, the methods always need to devise reducing numerical errors due to the projection between two discrete systems.
Moreover, the particle arrangements are free irrespective of the meshes, still causing un-equivalent computational stencils in space due to non-uniform distributions of the particles.

The aim of this study is to propose a novel approach for the use of compact stencils in particle-based meshless methods called the mesh-constrained discrete point (MCD) approach.
The MCD approach introduces a background mesh system in a domain and rigorously constrains the distribution of DPs in each mesh by solving a dynamic problem with nonlinear constraints.
This can avoid the heterogeneity of the DP distribution at the mesh-size level and impose compact stencils with a fixed degree of freedom for derivative evaluations.
A simple Cartesian mesh system is introduced as the background mesh that has the potential to make the computational efficiency of the present method competitive with that of conventional finite difference methods.
To ensure numerical accuracy, the MLS approximation is used for spatial derivatives (Tamai and Koshizuka, 2014; Matsunaga et al., 2020).
As compared with the existing particle-mesh hybrid methods, the proposed method is possible to solve the governing equations in a single discrete system consisting of DPs because our formulation does not require any projection between the DPs and background meshes.
In this study, a fundamental formulation is developed for arrangements of DPs in two dimensions and applied to unsteady Stokes flows.

This paper is organized as follows:
In Section 2, numerical formulations for two-dimensional Stokes equations are presented, including a flow solver based on a pressure projection method and discretization with MLS approximation.
In Section 3, the DP distribution algorithm is described.
In Section 4, numerical tests are presented for the evaluation of the DP distribution and flow problems with co-axial and eccentric circles.
The results are investigated in terms of spatial convergence accuracy and validity with respect to reference solutions.
Some concluding remarks are presented in Section 5.

\section{Flow solver} \label{flowSolver}
\subsection{Governing equations} \label{governingEquation}
In this study, the creeping motion of an incompressible and Newtonian viscous fluid is considered. The governing equations are given by unsteady Stokes equations:

\begin{equation}
  \nabla \cdot \textbf{v} = 0,
  \label{eq_Continuity}
\end{equation}

\begin{equation}
  \frac{\partial \textbf{v}}{\partial t} = - \nabla P + \nu \nabla^2 \textbf{v},
  \label{eq_Stokes}
\end{equation}
where $t>0$ is time, $\textbf{x}\in\mathbb{R}^d$ denotes the field positions in domain $\Omega$, $\textbf{v}(\textbf{x},t)\in\mathbb{R}^d$ is the fluid velocity, $P(\textbf{x},t)$ is the density-scaled pressure, $\nu$ is the kinematic viscosity, and $\partial / \partial t$ and $\nabla = \partial / \partial \textbf{x}$ are the partial derivatives of $t$ and $\textbf{x}$, respectively.
In this study, two-dimensional space $d=2$ is considered; that is, $\textbf{x}=(x,y)$ and $\textbf{v}=(u,v)$.

\subsection{Solution method}
Eqs. (\ref{eq_Continuity}) and (\ref{eq_Stokes}) are solved by the pressure projection method.
By applying first-order temporal discretization, the following semi-discrete system is obtained:

\begin{equation}
  \textbf{v}^* = \textbf{v}^k + {\Delta t}\nu \nabla^2 \textbf{v}^k,
  \label{eq_fStep1}
\end{equation}

\begin{equation}
  \nabla^2 P^{k+1} = \frac{1}{\Delta t} \nabla \cdot \textbf{v}^*,
  \label{eq_fStep2}
\end{equation}

\begin{equation}
  \textbf{v}^{k+1} = \textbf{v}^* - \Delta t \nabla P^{k+1},
  \label{eq_fStep3}
\end{equation}
where superscript $k$ denotes the time step, $\Delta t$ is the time interval, and $\textbf{v}^*$ is the intermediate velocity. 
Because steady-state fluid motion is the focus of this study, the above system is successively solved until the numerical solution converges.

In this study, the no-slip condition is prescribed on the boundary $\Gamma$ that enforces the Dirichlet boundary condition for the velocity as ${\bf v}=0$ on $\Gamma$. A further requirement of the projection method is the introduction of a boundary condition for the pressure Poisson equation (\ref{eq_fStep2}). The Neumann boundary condition is imposed as $\partial P/\partial n=\nabla{P}\cdot\textbf{n}=0$, where $\textbf{n}$ is the unit normal vector on $\Gamma$.

\subsection{Spatial discretization}
\label{discretization}
Discrete quantities are defined on distributed points in domain $\Omega$ and boundary $\Gamma$ that are constrained on background meshes, and hence the MLS method is applied for the discretization of spatial derivatives.
Although the basic formulation underlies the well-verified approach (Tamai and Koshizuka, 2014; Matsunaga et al., 2020), a unified derivation is introduced to evaluate derivatives for different quantities.

\subsubsection{General formulation}
Let $\Gamma_{\rm D}$ denote the Dirichlet boundary, $\Gamma_{\rm N}$ the Neumann boundary, and $\Omega_{\rm I}=\Omega \setminus (\Gamma_{\rm D} \cup \Gamma_{\rm N})$ the inner domain.
An arbitrary quantity defined as $\phi(\textbf{x})$ for $\textbf{x} \in \Omega$ is approximated in compact support domain $D_c = D(\textbf{x}_c)$ for arbitrary point $\textbf{x}_c$ as $\phi(\textbf{x}) \approx \Phi(\textbf{x})$. The local coordinate system in $D_c$ is introduced as

\begin{equation}
  \textbf{X} 
  = \frac{\textbf{x}-\textbf{x}_c}{r_{\rm s}}
  = \left[ X,~ Y \right]^{\top},
  \label{eq_scalingPosition}
\end{equation}
with scaling parameter $r_{\rm s}>0$. A polynomial approximation of $\phi(\textbf{x})$ in $D_c$ can be represented as

\begin{equation}
  \Phi(\textbf{X}) 
  = \textbf{p}(\textbf{X}) \cdot \tilde{\bm{\Phi}} + \phi_c, 
  \label{eq_Phi}
\end{equation}
where $\phi_c=\phi(\textbf{x}_c)$ is the quantity at $\textbf{x}_c$, $\textbf{p}(\textbf{X})$ is the polynomial basis vector, and $\tilde{\bm{\Phi}}$ is the modal vector (or polynomial coefficients).
In this study, the second-order polynomial is used as follows:

\begin{equation}
  \textbf{p}(\textbf{X}) = \left[ X,~ Y,~ X^2,~ X Y,~ Y^2 \right]^{\top},
  \label{eq_basis}
\end{equation}

\begin{equation}
  \tilde{\bm{\Phi}} = \left[ \tilde{\Phi}_{10},~ \tilde{\Phi}_{01},~ \tilde{\Phi}_{20},~ \tilde{\Phi}_{11},~ \tilde{\Phi}_{02} \right]^{\top}.
  \label{eq_tildePhi}
\end{equation}

The spatial derivative in the normal direction of $\Phi(\textbf{X})$ is represented using the unit normal vector $\textbf{n}=(n_x, n_y)$ on $\Gamma_{\rm N}$ as

\begin{equation}
  \frac{\partial \Phi}{\partial n} 
  = \textbf{n}\cdot\nabla{\Phi}
  = \left( 
      n_x \frac{\partial X}{\partial x} \frac{\partial \textbf{p}}{\partial X}
    + n_y \frac{\partial X}{\partial y} \frac{\partial \textbf{p}}{\partial Y}
    \right) \cdot \tilde{\bm{\Phi}}
  = \frac{1}{r_{\rm s}} \left( 
      n_x \frac{\partial \textbf{p}}{\partial X}
    + n_y \frac{\partial \textbf{p}}{\partial Y}
    \right) \cdot \tilde{\bm{\Phi}}
  = \frac{1}{r_{\rm s}} \textbf{p}^{\rm N}(\textbf{X}) \cdot \tilde{\bm{\Phi}},
  \label{eq_normalDerivative_Phi}
\end{equation}
where $\textbf{p}^{\rm N}(\textbf{X})$ is the polynomial basis vector with respect to the normal derivative:

\begin{equation}
  \textbf{p}^{\rm N}(\textbf{X}) 
  = \left[ n_x,~ n_y,~ 2 n_x X,~ n_x Y + n_y X,~ 2 n_y Y \right]^{\top}.
  \label{eq_basis_neumann}
\end{equation}

Assuming the discrete quantities $\phi_i = \phi(\textbf{x}_i)$ for point $i \in [1,n_{\rm DP}]$, where $n_{\rm DP}$ is the number of DPs, the polynomial function $\Phi(\textbf{X})$ is constructed through a minimization problem for objective function $J$:

\begin{equation}
  J = \frac{1}{2} \!\!\! \sum_{
      {\fontsize{7pt}{0mm}\selectfont
        \begin{array}{l}
          j \! \in \! \Lambda_c
        \end{array}}
      }{\!\!\!
      w_j \left( \textbf{p}_j \cdot \tilde{\bm{\Phi}} +\phi_c -\phi_j \right)^2}
    + \frac{1}{2} \!\!\! \sum_{
      {\fontsize{7pt}{0mm}\selectfont
        \begin{array}{l}
          j \! \in \! \Lambda^{\rm N}_c
        \end{array}}
      }{\!\!\!
      w_j \left\{ r_{\rm s} \left( \frac{1}{r_{\rm s}} \textbf{p}^{\rm N}_j \cdot \tilde{\bm{\Phi}} -f_{j} \right) \right\}^{2}}
    + \chi_c \lambda (\phi_c - g),
  \label{eq_objFunc}
\end{equation}
where
\begin{equation}
  \Lambda_c = \left\{ 
    i \in [1,n_{\rm DP}] \mid \textbf{x}_i \in D_c, \textbf{x}_i \in \Omega_{\rm I} \cup \Gamma_{\rm D} \right\},~~
  \Lambda^{\rm N}_c = \left\{ 
    i \in [1,n_{\rm DP}] \mid \textbf{x}_i \in D_c, \textbf{x}_i \in \Gamma_{\rm N} \right\},
  \label{eq_set}
\end{equation}
and $\textbf{p}_j = \textbf{p} \left( \textbf{X}_j \right)$, $\textbf{p}^{\rm N}_j = \textbf{p}^{\rm N} \left( \textbf{X}_j \right)$, $f_{j} = f(\textbf{x}_j)$ with function $f(\textbf{x})$ for the Neumann boundary condition, and $w_j=w \left( ||\textbf{x}_j - \textbf{x}_c|| \right)$ is the arbitrary weight.
The last term in Eq. (\ref{eq_objFunc}) is the constraint that enforces $\phi_c=g$, where $\lambda$ is the Lagrange multiplier, and $\chi_c$ is the characteristic function that is 1 or 0 according to whether the constraint is enforced or not, respectively.

To minimize $J$ with respect to $\phi_c$, $\tilde{\bm{\Phi}}$, and $\lambda$, the following stationary conditions are derived:

\begin{equation}
  \begin{array}{l}
    \displaystyle ~~~~~
      \frac{\partial J}{\partial \phi_c} = 0,~~
      \frac{\partial J}{\partial \tilde{\bm{\Phi}}} = \bm{0},~~
      \frac{\partial J}{\partial \lambda} = 0,
    \quad \Rightarrow \quad
    \left\{
      \begin{array}{l}
        \displaystyle 
          a \phi_c + \textbf{b} \cdot \tilde{\bm{\Phi}} + \chi_c \lambda
          = c, \\
        \displaystyle 
          \textbf{b} \phi_c + \left(\textbf{L}+\textbf{L}^{\rm N}\right) \tilde{\bm{\Phi}}
          = \textbf{d} + \textbf{d}^{\rm N}, \\
        \displaystyle 
          \chi_c \phi_c
          = \chi_c g,
      \end{array}
    \right.
  \end{array}
    \label{eq_simulations}
\end{equation}
where

\begin{equation}
  a
  = \!\!\! \sum_{
    {\fontsize{7pt}{0mm}\selectfont
      \begin{array}{l}
        j \! \in \! \Lambda_{c}
      \end{array}}
    }{\!\!\!
    w_{j} }, \quad
  \label{eq_a}
\end{equation}

\begin{equation}
  \textbf{b}
  = \!\!\! \sum_{
    {\fontsize{7pt}{0mm}\selectfont
      \begin{array}{l}
        j \! \in \! \Lambda_{c}
      \end{array}}
    }{\!\!\!
    w_{j} \textbf{p}_{j} }, \quad
  \label{eq_b}
\end{equation}

\begin{equation}
  c
  = \!\!\! \sum_{
    {\fontsize{7pt}{0mm}\selectfont
      \begin{array}{l}
        j \! \in \! \Lambda_{c}
      \end{array}}
    }{\!\!\!
    w_{j} \phi_{j} }
  \label{eq_c}
\end{equation}

\begin{equation}
  \textbf{d}
  = \!\!\! \sum_{
    {\fontsize{7pt}{0mm}\selectfont
      \begin{array}{l}
        j \! \in \! \Lambda_{c}
      \end{array}}
    }{\!\!\!
    w_{j} \phi_{j} \textbf{p}_{j} }, \ \
  \textbf{d}^{\rm N}
  = \!\!\! \sum_{
    {\fontsize{7pt}{0mm}\selectfont
      \begin{array}{l}
        j \! \in \! \Lambda^{\rm N}_{c}
      \end{array}}
    }{\!\!\!
    r_{\rm s} w_{j} f_{j} \textbf{p}^{\rm N}_{j} }. 
  \label{eq_d}
\end{equation}

\begin{equation}
  \textbf{L}
  = \!\!\! \sum_{
    {\fontsize{7pt}{0mm}\selectfont
      \begin{array}{l}
        j \! \in \! \Lambda_{c}
      \end{array}}
    }{\!\!\!
    w_{j} \textbf{p}_{j}\textbf{p}_{j}^{\top} }, \ \
  \textbf{L}^{\rm N}
   =
   \!\!\! \sum_{
    {\fontsize{7pt}{0mm}\selectfont
      \begin{array}{l}
        j \! \in \! \Lambda^{\rm N}_{c}
      \end{array}}
    }{\!\!\!
    w_{j} \textbf{p}^{\rm N}_{j}(\textbf{p}^{\rm N}_{j})^{\top} }.
  \label{eq_L}
\end{equation}
By eliminating $\phi_c$, Eq. (\ref{eq_simulations}) can be written as
\begin{equation}
  \textbf{M} \tilde{\bm{\Phi}} = \textbf{e},
  \label{eq_simulations2}
\end{equation}
where

\begin{equation}
  \textbf{M} = \textbf{L} + \textbf{L}^{\rm N},
  \label{eq_M}
\end{equation}

\begin{equation}
  \textbf{e} = \textbf{d} + \textbf{d}^{\rm N} - g\textbf{b},
  \label{eq_e}
\end{equation}
for $\chi_c=1$, and 

\begin{equation}
  \textbf{M} = \textbf{L} + \textbf{L}^{\rm N} -\frac{1}{a} \textbf{b}\textbf{b}^{\top},
  \label{eq_M2}
\end{equation}

\begin{equation}
  \textbf{e} = \textbf{d} + \textbf{d}^{\rm N} - \frac{c}{a} \textbf{b}.
  \label{eq_e2}
\end{equation}
for $\chi_c=0$.
When the moment matrix $\textbf{M}$ is non-singular, the modal components of $\tilde{\bm{\Phi}}$ are obtained as $\tilde{\bm{\Phi}}=\textbf{M}^{-1}\textbf{e}$.
Additionally, quantity $\phi_c$ is evaluated as $\phi_c =g$ for $\chi_c=1$ or $\phi_c = (c-\textbf{b}\cdot\tilde{\bm{\Phi}})/a$ for $\chi_c=0$.
Thus, the $k$-th order spatial derivatives (up to $k=2$ in this study) at $\textbf{x}_c$ are obtained by differentiating polynomial $\Phi(\textbf{x})$ and written as

\begin{equation}
  \left. \bm{\partial} \phi \right|_{c}
  = \textbf{H} \tilde{\bm{\Phi}},
  \label{eq_derivativeVector}
\end{equation}

\begin{equation}
  \bm{\partial} 
  = \left[ \displaystyle
  \frac{\partial}{\partial x},~ 
  \frac{\partial}{\partial y},~ 
  \frac{\partial^2}{\partial x^2},~ 
  \frac{\partial^2}{\partial x \partial y},~ 
  \frac{\partial^2}{\partial y^2}
 \right]^{\top}, 
 \label{eq_differentialOperatorVector}
\end{equation}

\begin{equation}
  \textbf{H}
  = {\rm diag} \left( 
      ~ \frac{1}{r_{\rm s}},
      ~ \frac{1}{r_{\rm s}},
      ~ \frac{2}{r^{2}_{\rm s}},
      ~ \frac{1}{r^{2}_{\rm s}},
      ~ \frac{2}{r^{2}_{\rm s}}
    \right), 
  \label{eq_Hrs}
\end{equation}
where $\textbf{H}$ is the scaling diagonal matrix.

\subsubsection{Evaluation at  DP $i$}
Eqs. (\ref{eq_fStep1}), (\ref{eq_fStep2}), and (\ref{eq_fStep3}) are solved in a strong form for the discrete quantities at $\textbf{x}_i$ ($1 \leq i \leq n_{\rm DP})$, and thus the evaluation position for the spatial derivatives is equivalent to the DPs; that is, $\textbf{x}_c = \textbf{x}_i$.
When $\chi_c=1$ and $g = \phi_c = \phi_i$ are imposed, the above-mentioned MLS reconstruction is performed for DP $i$.
The notation $\cdot|_{i}$ is used to describe the quantity of $i$.

The vector $\textbf{e}$ in Eq. (\ref{eq_e}) is divided into function values and the Neumann boundary term as follows:

\begin{equation}
  \textbf{e}|_i
  = \!\!\! \sum_{
    {\fontsize{7pt}{0mm}\selectfont
      \begin{array}{l}
        j \! \in \! \Lambda_{i}
      \end{array}}
    }{\!\!\!
    \textbf{f}_{j}|_{i} \, (\phi_{j}-\phi_{i}) + \textbf{g}|_{i} }, 
  \label{eq_divide_e}
\end{equation}
where
\begin{equation}
  \textbf{f}_{j}|_{i} = w_{j}|_{i} \, \textbf{p}_{j}|_{i},
  \quad
  \textbf{g}|_{i} = \textbf{d}^{\rm N}|_{i}.
  \label{eq_f_g}
\end{equation}
Consequently, the derivative vector of $\phi$ at $\textbf{x}_{i}$ (Eq. (\ref{eq_derivativeVector})) can be rewritten as

\begin{equation}
  \left. \bm{\partial} \phi \right|_{i}
  = \textbf{H} \textbf{M}|_{i}^{-1} \left(
      \!\!\! \sum_{
      {\fontsize{7pt}{0mm}\selectfont
        \begin{array}{l}
          j \! \in \! \Lambda_{i}
        \end{array}}
      }{\!\!\!
      \textbf{f}_{j}|_{i} \, (\phi_{j}-\phi_{i}) + \textbf{g}|_{i} }
    \right)
  = \!\!\! \sum_{ 
    {\fontsize{7pt}{0mm}\selectfont
      \begin{array}{l}
        j \! \in \! \Lambda_{i}
      \end{array}}
    }{\!\!\!
    \hat{\textbf{f}}_{ij} \phi_{j}} 
    - \hat{\textbf{h}}_{i} \phi_{i}
    + \hat{\textbf{g}}_{i},
  \label{eq_derivativeVector2}
\end{equation}
where $\hat{\textbf{f}}_{ij} = \textbf{H} \textbf{M}|_{i}^{-1} \textbf{f}_{j}|_{i}$, $\hat{\textbf{g}}_{i} = \textbf{H} \textbf{M}|_{i}^{-1} \textbf{g}|_{i}$, and $\hat{\textbf{h}}_{i} = \sum_{j \in \Lambda_i} \hat{\textbf{f}}_{ij}$.

This form can be used to derive a linear system, such as the Poisson equation.
For instance, the Laplacian of $\phi$ at $\textbf{x}_{i}$ can be written as

\begin{equation}
  \nabla^2 \phi|_{i} 
  = \left. \bm{\partial} \phi \right|_{i} 
  \cdot \left[ 0,~ 0,~ 1,~ 0,~ 1 \right]^{\top}
  = \!\!\! \sum_{ 
    {\fontsize{7pt}{0mm}\selectfont
      \begin{array}{l}
        j \! \in \! \Lambda_{i}
      \end{array}}
    }{\!\!\!
    C_{ij} \phi_{j} } + r_i, 
    \label{eq_laplacian_q}
\end{equation}

\begin{equation}
  C_{ij} = \left(\hat{\textbf{f}}_{ij} - \delta_{ij}\hat{\textbf{h}}_{i} \right) \cdot \left[ 0,~ 0,~ 1,~ 0,~ 1 \right]^{\top},~~
  r_i = \hat{\textbf{g}}_{i} \cdot \left[ 0,~ 0,~ 1,~ 0,~ 1 \right]^{\top}.
  \label{eq_Cij}
\end{equation}

\section{MCD approach} \label{MCDmethod}

\subsection{Overview} \label{overview}
In the present mesh-constrained approach, DPs are linked to an arbitrary (background) mesh system that encompasses analysis domain $\Omega$.
Although the background meshes are secondarily used in conventional particle methods as bucket-based data management for particle positions, the present MCD approach primarily uses the background meshes for the definition of the DPs and core formulation in the discretization.
This yields more compactness of calculation stencils than that of conventional meshless and particle methods, and attains high computational efficiency and load balance in parallel computation that is competitive with mesh-based approaches.

The Cartesian mesh system is used for the background meshes.
The DPs are defined so that each of them belongs to a unique (or non-overlapped) mesh, and the positions are determined to lie in the inner domain of $\Omega$ or on the boundary $\Gamma$, depending on the mesh configuration.

\subsection{Background meshes and initial/temporal points}
\label{makeGrid}

The center positions of the Cartesian meshes, $\textbf{x}_{i,j} = (x_i, y_j)$, are defined with respect to the $x$ and $y$ directions as

\begin{equation}
  x_i = \left( i - \frac{1}{2} \right) \Delta x + x_0 ~~ \left( i = 1,~ 2,~,... ,~ N_x \right),
  \label{eq_xiDP}
\end{equation}

\begin{equation}
  y_j = \left( j - \frac{1}{2} \right) \Delta y + y_0 ~~ \left( j = 1,~ 2,~,... ,~ N_y \right),
  \label{eq_yjDP}
\end{equation}
where $N_x$, $N_y$ are the numbers of meshes, $x_0$, $y_0$ are the coordinates of the meshes, and $\Delta x$ $\Delta y$ are the mesh widths, which are $h = \Delta x = \Delta y$ in this study. 

The temporal (discrete) points are initially located on ${\bf x}_{i, j}$, where the representative distance of the DPs is defined as $l_0 = h$.


\subsection{Representation of arbitrary boundary shapes using the signed distance function  (SDF)}
\label{makeSDF}

To represent arbitrary boundary shapes, the signed distance function is introduced, $\psi({\bf x})$, which is $\psi>0$ for the dmain inside the domain ($\textbf{x} \in \Omega_{\rm I}$), $\psi<0$ for the domain outside the domain ($\textbf{x} \notin \Omega$), and $\psi=0$ for the boundary ($\textbf{x} \in \Gamma$).
The SDF is discretely given as $\psi_{i,j}$ $(i = 1, 2,... , N^{\rm SDF}_{x}$, $j = 1, 2,... , N^{\rm SDF}_{y})$, defined on a uniform Cartesian grid system $(x^{\rm SDF}_{i}, y^{\rm SDF}_{j})$, with number of grids $N^{\rm SDF}_{x}$, $N^{\rm SDF}_{y}$, origin $x^{\rm SDF}_{0}$, $y^{\rm SDF}_{0}$, and grid widths $\Delta x^{\rm SDF}$, $\Delta y^{\rm SDF}$ in the $x$ and $y$ directions, respectively.
Through a numerical interpolation from the regularly aligned $\psi_{i,j}$, it is easy to evaluate the SDF value $\psi_c$ and its derivatives $\nabla \psi_c$ at an arbitrary position $\textbf{x}_c$.
In this study, MLS interpolation based on Eq. (\ref{eq_simulations2}) with Eqs. (\ref{eq_M2}) and (\ref{eq_e2}) is applied.
A radial compact support domain $D_c^{\rm SDF} = \{\textbf{x} \mid ||\textbf{x}-\textbf{x}_c|| \leq r_{\rm e}^{\rm SDF}\}$ is set, and weight $w_j^{\rm SDF} = w^{\rm SDF} \left( r_{\rm e}^{\rm SDF}; ||\textbf{x}_j - \textbf{x}_c|| \right)$, with a weight function:
\begin{equation}
  w^{\rm SDF}(r_{\rm e};r)
  = \left\{ \begin{array}{cl}
      \displaystyle
        \frac{1}{2} \left( 1 + \cos{ \frac{\pi r}{r_{\rm e}} } \right), & \textrm{for} \ r \leq r_{\rm e}, \\
      0, & \textrm{otherwise},
    \end{array} 
  \right.
  \label{eq_weight}
\end{equation}
where $r_{\rm e}^{\rm SDF}$ is the influence radius for the MLS reconstruction.

Note that, in the case that the DPs are not moved during fluid calculations assumed in this study, the evaluation for SDFs becomes pre-process before fluid simulations, and thus the computational efficiency does not matter.
Therefore, the radial compact support $D_{\rm e}^{\rm SDF}$ and weight $w^{\rm SDF}(r_e;r)$, which has been well validated, were applied.

\subsection{Assignment of masks for temporal points}
\label{maskDP}

\begin{figure}[H]
  \begin{center}
    \includegraphics[width=90mm]{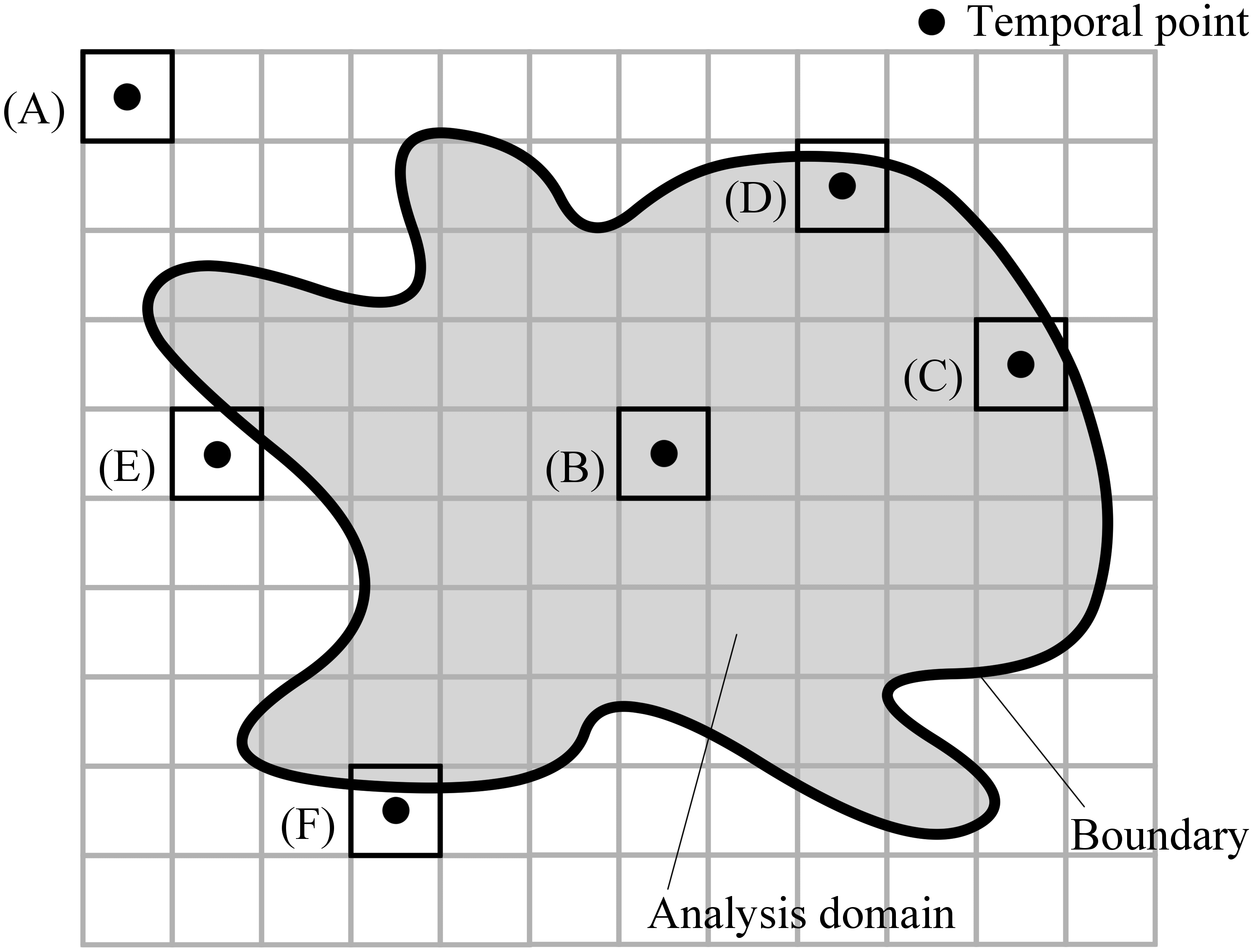}
    \caption{Typical cases for the masks of temporal (or discrete) points, which are initially located at the center position of the background meshes. When the temporal point is far from the boundary, where distance $d$ is larger than $l_0/2$, the mask is set to 0 for the outside the boundary (A, E) or 1 for the inside the boundary (B, C). When the temporal points are near the boundary ($d \leq l_0/2$), the masks are set to 2 (D, F).}
    \label{fig_maskDP}
  \end{center}
\end{figure}

The temporal (and discrete) points are labeled with ``mask" depending on the configuration of the background meshes, which indicates that position $\textbf{x}_i$ is constrained inside the domain ($\textbf{x}_i\in\Omega_{\rm I}$), on the boundary ($\textbf{x}_i \in \Gamma$), and outside the domain ($\textbf{x}_i\notin\Omega$) (Fig. \ref{fig_maskDP}).

The definition of the mask is shown in Table \ref{table_maskValue}, and it is determined by SDFs $\psi_{i,j}$ for the temporal points:

\begin{equation}
  \textrm{mask}_{i,j}
  = \left\{ \begin{array}{ll}
    \displaystyle
    0, & \textrm{for} \ \psi_{i,j} < -\frac{l_0}{2}, \\
    1, & \textrm{for} \ \psi_{i,j} > \frac{l_0}{2}, \\
    2,3,\cdots, & \textrm{otherwise}.
  \end{array} 
\right.
  \label{mask}
\end{equation}
Note that the boundary mask ($\geq 2$) is used to identify multiple boundaries. The temporal points with mask $\neq$ 0 are called DPs, which are used for the calculation.

\begin{table}[H]
  \begin{center}
    \caption{ Definition of the mask.}
      \begin{tabular}{ll} \hline
        type & mask \\ \hline
        Outside & 0 \\
        Inside & 1 \\
        Boundary & 2,3,... \\ \hline
      \end{tabular}
    \label{table_maskValue}
  \end{center}
\end{table}

\subsection{Arrangement of the DPs}
\label{equalPlacement}
To arrange the DPs, an equilibrium problem is solved for their position.
The solution is obtained by solving a dynamic problem for their motion under two nonlinear constraints: (i) relocate the points with mask $\geq$ 2 to the boundary surface and (ii) restrict them to exceed each background grid. 
This enables the DPs to be distributed without an extremely coarse or fine arrangement in space, located on the boundary surface, and maintained in the initially assigned background meshes.

\begin{figure}[H] 
  \begin{center}
    \includegraphics[width=140mm]{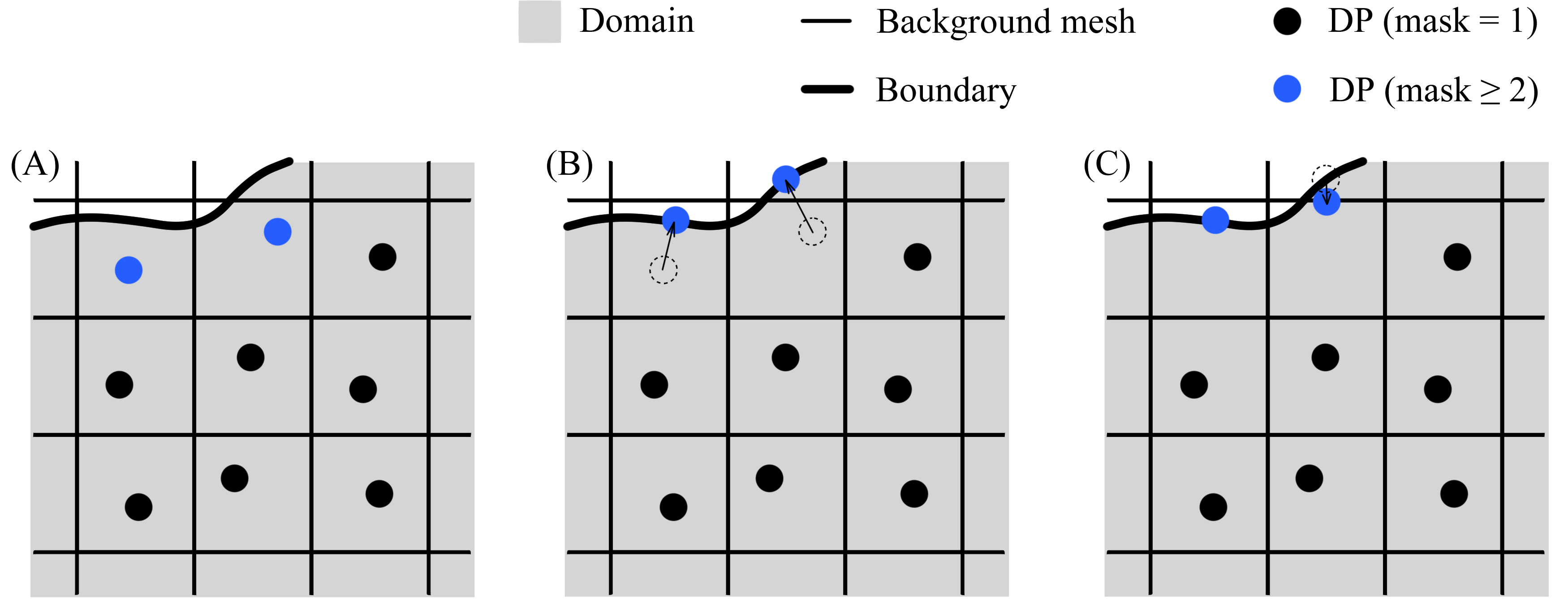}
    \caption{Single sequence in the step-by-step approach for the arrangement of DPs. (A) The DPs are moved by solving the dynamic problem considering an interaction with surrounding DPs (included in $3 \times 3$ background grids). (B) The DPs with mask $\geq$ 2 are relocated to the boundary surface. (C) The DPs that exceed the initially assigned respective meshes are pulled back to the grid boundary. This sequence is repeated until the distribution of DPs reaches equilibrium.}
    \label{fig_moveDP}
  \end{center}
\end{figure}

Solving the dynamic problem under the above-mentioned nonlinear constraints is typically complicated; hence, a step-by-step formulation is applied to obtain the equilibrium position (Fig. \ref{fig_moveDP}). The following sequence is repeated until the DP positions converge.

\subsubsection*{Repositioning based on a dynamic problem} \label{equilibriumProblem}
The following inertialess equation of motion is introduced for DP $i$:

\begin{equation}
  \gamma \textbf{v}_i + \textbf{F}_i = 0,
  \label{eq_geMCD1}
\end{equation}
where $\gamma$ is the damping coefficient, $\textbf{v}_i = d \textbf{x}_i / dt$ is the velocity, and $\textbf{F}_i$ is the resultant force caused by interactive forces on neighboring points $j$, $\textbf{F}_{ij}$:

\begin{equation}
  \textbf{F}_i 
  = \left\{ \begin{array}{cl}
      \displaystyle
        \!\!\!\! \sum_{
          {\fontsize{7pt}{0mm}\selectfont \begin{array}{c}
            j \in \Lambda^{\rm I}_i \cup \Lambda^{\Gamma}_i
          \end{array}}
        }{\!\!\!\! \textbf{F}_{ij} }, & \left( i \in \Lambda^{\rm I}_i \right), \\[7mm]
      \displaystyle
        \!\!\!\! \sum_{
          {\fontsize{7pt}{0mm}\selectfont \begin{array}{c}
            j \in \Lambda^{\Gamma}_i
          \end{array}}
        }{\!\!\!\! \textbf{F}_{ij} }, & \left( i \in \Lambda^{\Gamma}_i \right),
    \end{array}
  \right.
  \label{eq_geMCD1_kp}
\end{equation}
where
\begin{equation}
  \Lambda^{\rm I}_i 
  = \left\{ j \in [1,n_{\rm DP}] \mid j \neq i, \, 
  \textbf{x}_j \in D_i^{\rm DP}, \, 
  \textbf{x}_j \in \Omega_{\rm I} \right\},
  \label{eq_Lambda_i}
\end{equation}
\begin{equation}
  \Lambda^{\Gamma}_i 
  = \left\{ j \in [1,n_{\rm DP}] \mid j \neq i, \, 
  \textbf{x}_j \in D_i^{\rm DP}, \,
  \textbf{x}_j \in \Gamma \right\}
  \label{eq_Lambda_g}
\end{equation}
are the sets of indices for the domain inside and the boundary, respectively (i.e., compact sets of neighboring DPs at mask = 1 or $\geq 2$ for $i$), where $D_i^{\rm DP}$ is the compact support for $i$ that depends on the background meshes. 
In this study, the interactive force $\textbf{F}_{ij}$ is given by

\begin{equation}
  \textbf{F}_{ij}
  = F_0 \hat{F}_{ij} \hat{\textbf{x}}_{ij},
  \label{eq_forceMCD}
\end{equation}
where $F_0$ is the constant force strength, $\hat{\textbf{x}}_{ij}$ is the unit vector of the direction vector $\textbf{x}_{ij} = \textbf{x}_{j}-\textbf{x}_{i}$, and $\hat{F}_{ij} = G(||\textbf{x}_{ij}||/l_0)$ denotes the adimensional force function, assuming linear short-range repulsion:

\begin{equation}
  G(r')
  = \left\{ \begin{array}{cl} \displaystyle
    1 - r', & \textrm{for} \ r' \leq 1, \\[3mm]
    0, & \textrm{otherwise}.
    \end{array} \right.
  \label{eq_Fij_xij}
\end{equation}
Because the influence radius of repulsion between the DPs is set to $l_{0}$, the compact support $D_i^{\rm DP}$ is sufficient to set $3 \times 3$ background meshes, where the center mesh includes point $i$.

By applying the explicit Euler method with time interval $\Delta \tau$, an instantaneous position $\textbf{x}^*$ is obtained as

\begin{equation}
  \textbf{x}^{*}_{i} 
  = \textbf{x}^{k}_{i} + {\Delta \tau} \textbf{v}^{k}_{i}
  = \textbf{x}^{k}_{i} - \frac{\Delta \tau}{\gamma} \textbf{F}^{k}_i,
  \label{eq_position}
\end{equation}
where superscript $k$ denotes the $k$-th time step.
Suppose $l_0$ and $\Delta \tau$ are the characteristic length and time. Eq. (\ref{eq_position}) is then replaced by the following dimensionless form:

\begin{equation}
  \hat{\textbf{x}}^{*}_{i} 
  = \hat{\textbf{x}}^{k}_{i} + \hat{\textbf{v}}^{k}_{i}
  = \hat{\textbf{x}}^{k}_{i} - \kappa \hat{\textbf{F}}^{k}_i,
  \label{eq_position_nondim}
\end{equation}
with the following adimensional quantities:
\begin{equation}
  \hat{\textbf{x}}_{i} = \frac{\textbf{x}_{i}}{l_0}, \ \
  \hat{\textbf{v}}_{i} = \frac{\Delta \tau}{l_0} \textbf{v}_{i}, \ \
  \hat{\textbf{F}}_{i} = \frac{\textbf{F}_{i}}{F_0},
  \label{eq_adimension}
\end{equation}
and
\begin{equation}
  \kappa = \frac{\Delta \tau F_{0}}{\gamma l_0}.
  \label{eq_kappa}
\end{equation}

It is expected that the dimensionless quantity $\kappa$ results in similar behavior to DP motion in the dynamic problem and it is no longer necessary to set $F_{0}$, $\gamma$, and $\Delta \tau$ individually.
Additionally, solutions should be stable once $\kappa$ is adjusted to any spatial resolution $l_0$ (or $h$).

\subsubsection*{Relocation of DPs to the boundary surface} \label{constraint_boundary}
After the instantaneous (or intermediate) position $\textbf{x}^{*}_{i}$ is updated through Eq. (\ref{eq_position}), the next intermediate position $\textbf{x}^{**}_{i}$ to satisfy constraint (i) (Fig. \ref{fig_moveDP} (B)) is calculated by

\begin{equation}
  \textbf{x}^{**}_{i} = 
  \left\{ \begin{array}{cl}
    \textbf{x}^{*}_{i}, & \left( i \in \Lambda^{\rm I}_{i} \right), \\[2mm]
    \textbf{x}^{*}_{i} + \phi_{i} \hat{\textbf{d}}_i, & \left( i \in \Lambda^{\Gamma}_{i} \right),
  \end{array} \right.
  \label{eq_x**MCD}
\end{equation}
where $\phi_i$ is the SDF and $\hat{\textbf{d}}_i$ is the unit direction vector from point $i$ to the boundary surface given by $\hat{\textbf{d}}_i = - \nabla \phi_{i}/|\nabla \phi_{i}|$.
These values are evaluated by the MLS reconstruction at $\textbf{x}^{*}_i$ described in Section \ref{makeSDF}.

\subsubsection*{Relocation of DPs to the grid boundaries} \label{constraint_grid}
At this stage, the DPs are not guaranteed to be located in the initially assigned background meshes.
DPs that exceed the mesh region are finally pulled back to the mesh boundary.
Because of the simplicity of Cartesian meshes, the strategy is simply applied:

\begin{enumerate}
  \setlength{\leftskip}{11mm}
  \item[1.~~~] 
    If the DP position $\textbf{x}^{**}_i$ is inside the initially assigned mesh, $\textbf{x}^{k+1}_i = \textbf{x}^{**}_i$ is set;
  \item[2.~~~] 
    otherwise, if $\textbf{x}^{**}_i$ exceeds the grid boundary in the $\alpha$ direction, the $\alpha$ coordinate is altered to that of the nearest grid boundary and the position is set to $\textbf{x}^{k+1}_i$, where $\alpha=x,y$.
    Note that if the DP is outside the boundary for both the $x$ and $y$ directions, their $x$ and $y$ positions are pulled back to those of the grid boundaries for $x$ and $y$; that is, the DP is repositioned onto the nearby grid vertex.
\end{enumerate}

\begin{figure}[H]
  \begin{center}
    \includegraphics[width=170mm]{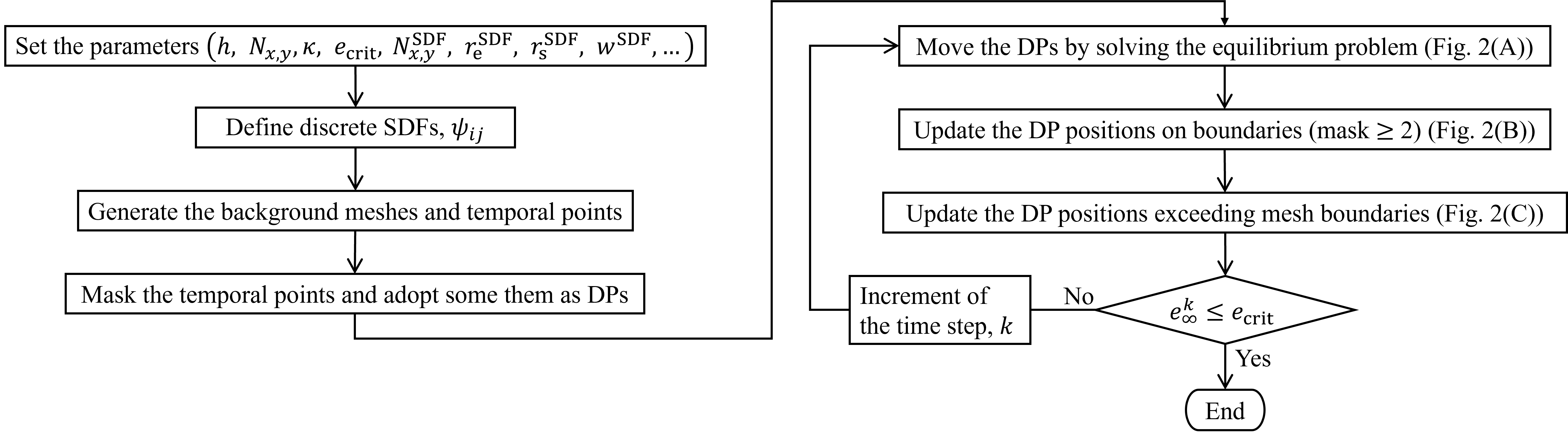}
    \caption{The flowchart for the DP arrangement, where $e^k_{\infty}$ denotes the $L_{\infty}$ norm of the relative displacement of DP and $e_{\rm crit}$ denotes the convergence criterion shown later.}
    \label{fig_shiftDP_Grid}
  \end{center}
\end{figure}

The algorithm for the DP arrangement is summarized as a flowchart in Fig. \ref{fig_shiftDP_Grid}.

\subsection{MLS reconstruction using the MCD approach} \label{MCD_MLS_relation}
Because of the present configuration of DPs, the compact support in the MLS reconstruction for the derivative evaluation can be determined based on the background meshes.
Because the quadratic polynomial is applied for the MLS reconstruction, compact support $D_i$ is set to $3 \times 3$ blocks of the background meshes from the center for the $i$-th DP, where the weight is set to 1; that is, $w_{j}|_{i}=1$. 

This enables the evaluation of the derivatives in governing equations using the same number of degrees of freedom ($3 \times 3 = 9$) for all DPs.
Moreover, the memory allocations for the unknowns defined at the DPs based on structured grids can be used, which results in efficient computational performance.

\section{Numerical tests}
\label{numericalTest}

\subsection{Distribution of discrete points} 
\label{distributionDP}

\begin{figure}[H]
  \begin{center}
    \includegraphics[width=65mm]{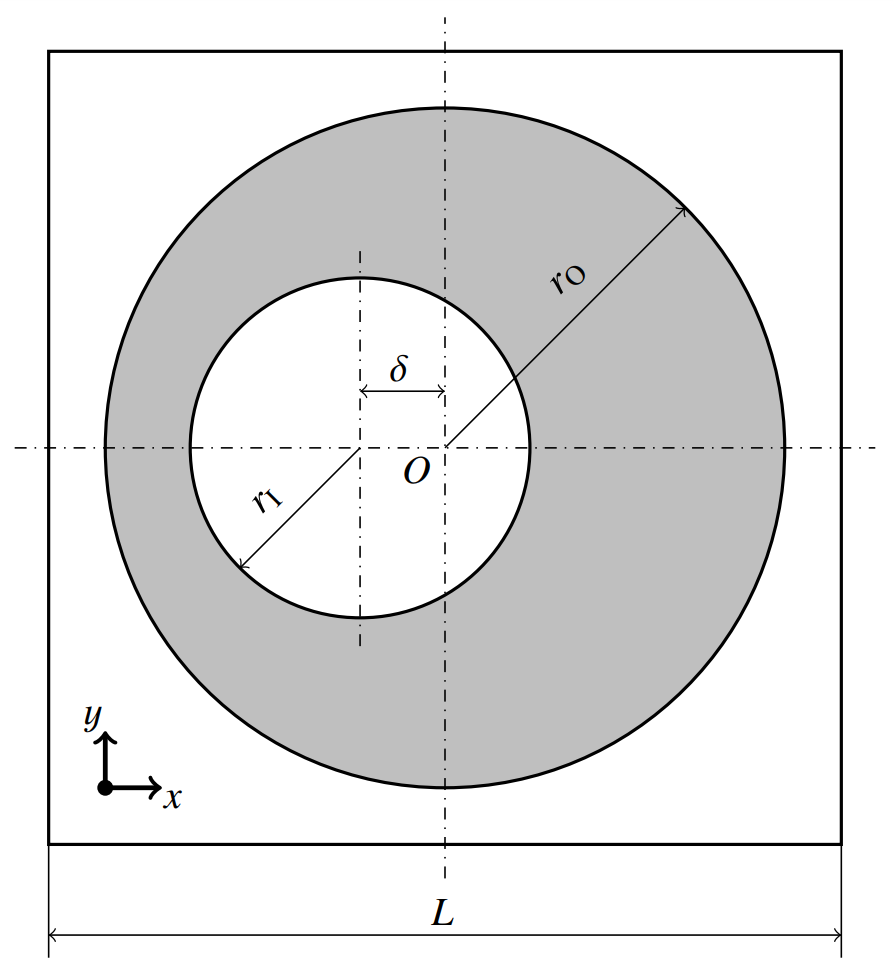}
    \caption{Illustration of the analysis configuration for two circular boundaries, which represent the co-axial circular case at $\delta = 0$ and the eccentric circular case for $0 < \delta < r_{\rm O} - r_{\rm I}$.}
    \label{fig_sketchDomain}
  \end{center}
\end{figure}

The distribution of DPs obtained by the proposed mesh-constrained approach for two circular boundaries is investigated.
As shown in Fig. \ref{fig_sketchDomain}, the inner and outer circles with radii of $r_{\rm I}$ and $r_{\rm O}$, respectively, are set to an $L \times L$ square domain, where $\delta$ is the eccentricity length in the negative $x$ direction.
Two cases are considered: co-axial circles and eccentric circles with the parameters shown in Table \ref{table_dimension}. 
The boundaries for the outer and inner circles are set to mask = 2 and mask = 3, respectively.
The background meshes are generated with $N \times N$ grids ($N_x=N_y=N$) with a grid width $h = l_0 = L/N$.
The grid width for the discrete SDFs is set to be the same as that of the background meshes for the DPs ($\Delta x^{\rm SDF} = \Delta y^{\rm SDF} = h$), and the domain length for the SDF is set to be larger than that of the DP by six grid widths. 
The MLS parameters used to evaluate $\phi_{i}$ and $\nabla \phi_{i}$ in Eq. (\ref{eq_x**MCD}) are set to $r^{\rm SDF}_{\rm e} = 3.1 h$ and $r^{\rm SDF}_{\rm s} = 0.7 h$.

\begin{table}[H]
  \begin{center}
    \caption{Geometric parameters in each analysis configuration.}
    \begin{tabular}{lcccc} \hline
      ~~ & $L$ & $r_{\rm I}$ & $r_{\rm O}$ & $\delta$ \\ \hline
      Co-axial circles & 5 & 1 & 2 & 0 \\
      Eccentric circles & 2.5 & 0.5 & 1 & 0.25 \\ \hline
    \end{tabular}
  \label{table_dimension}
  \end{center}
\end{table}

\begin{figure}[H] 
  \begin{center}
    \includegraphics[width=\linewidth]{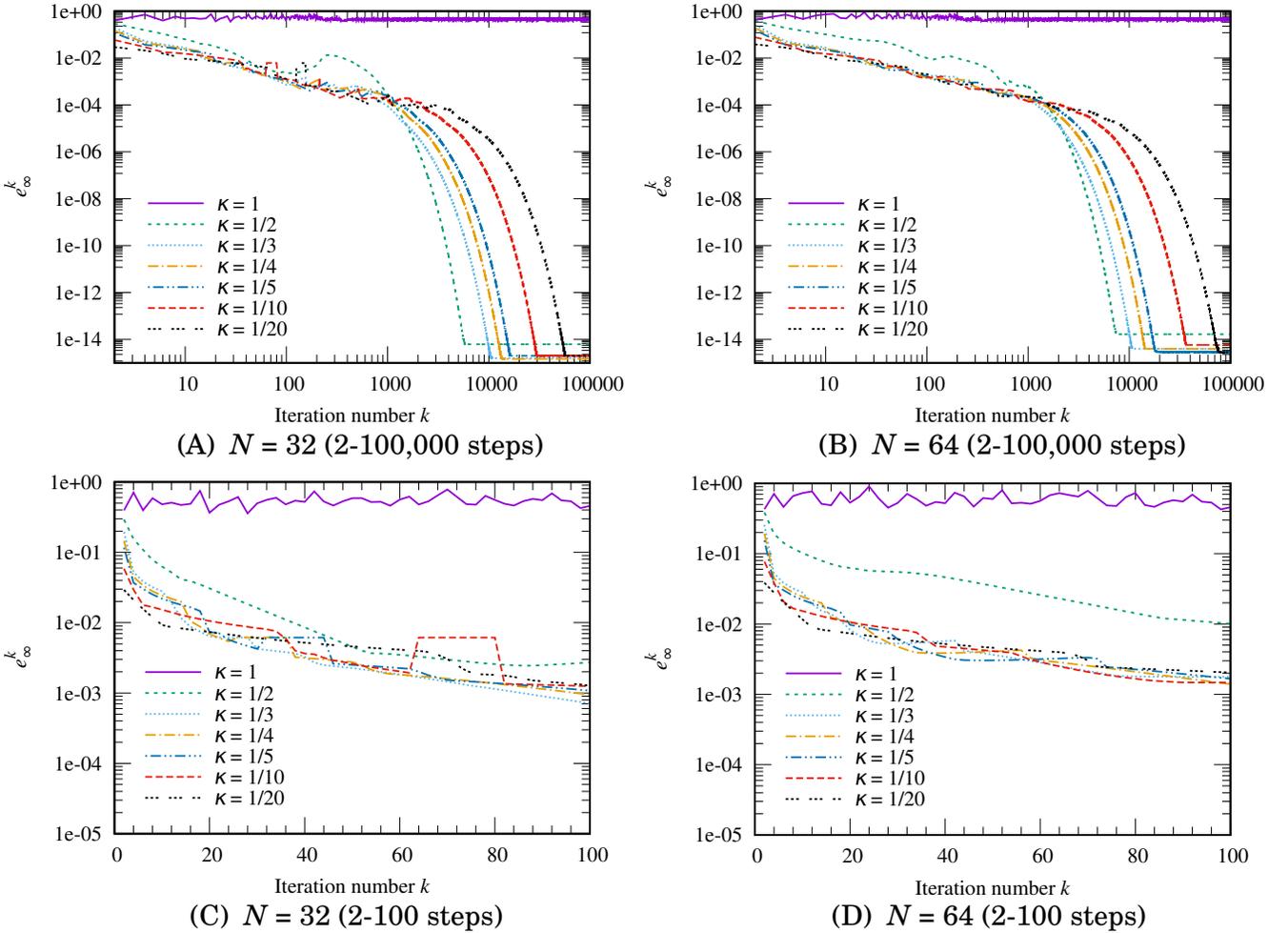}
    \caption{Comparisons of the convergence behavior of the relative displacement $e^{k}_{\infty}$ for different parameters of $\kappa$ at the background grids $N=32$ (A, C) and 64 (B, D), where (C), (D) show enlarged views for a range of iteration numbers for $k \leq 100$.}
    \label{fig_determineKappaLinf}
  \end{center}
\end{figure}

First, the effects of $\kappa$ given in Eq. (\ref{eq_kappa}) on the convergence behavior of DP motion is investigated.
The co-axial circular case is considered, with $N=32$ and 64.
The convergence behavior is quantified by the relative displacement:
\begin{equation}
  e^{k}_{\infty}
  = \frac{1}{l_{0}}
    \max_{i \in [1,N_{\rm DP}]}{ \left| \textbf{x}^{k}_{i} - \textbf{x}^{k-1}_{i} \right| }.
  \label{eq_Linf_MCD}
\end{equation}
Fig. \ref{fig_determineKappaLinf} shows comparisons of $e^{k}_{\infty}$ for different $\kappa$ and $N$.
The results demonstrate convergence behavior, except for $\kappa=1$, where $\kappa=1/2$ slowly converges and oscillates in early step iterations.
For $\kappa \leq 1/3$, similar convergence behavior is obtained in the early stage of iterations around $O(k) \leq 1000$, whereas convergence becomes more rapid in the latter stage.
Although the convergence behavior for $N=32$ is slightly earlier than that for $N=64$, its dependence on $\kappa$ is similar for both meshes.
It is expected that an increase of $\kappa$ would accelerate DP motion in the dynamic problem (\ref{eq_geMCD1}). 
The results indicate that the latter-stage differences in convergence among $\kappa$ originate from the dynamic problem, whereas the early-stage behavior is dominantly attributed to the kinematic constraints on repositioning to the boundary surface and grid boundary that are independent of $\kappa$.
The results confirm that the proposed algorithm provides stable and good converged results when $\kappa$ is appropriately chosen.
Hereafter, $\kappa = 1/3$ is adopted in numerical tests.

\begin{figure}[H] 
  \begin{center}
    \includegraphics[width=100mm]{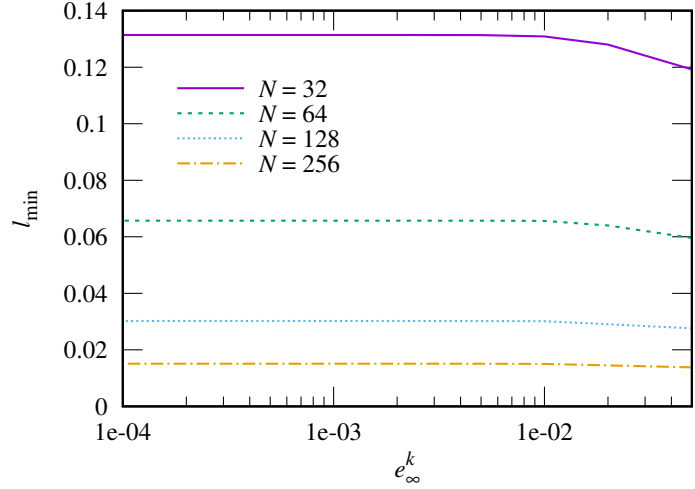} 
    \caption{Relationships between relative displacement $e^{k}_{\infty}$ and minimum distance between DPs $l_{\rm min}$ for different background meshes with $N$.}
    \label{fig_compareTolerance_l_min}
  \end{center}
\end{figure}

\begin{figure}[H] 
  \begin{center}
    \includegraphics[width=100mm]{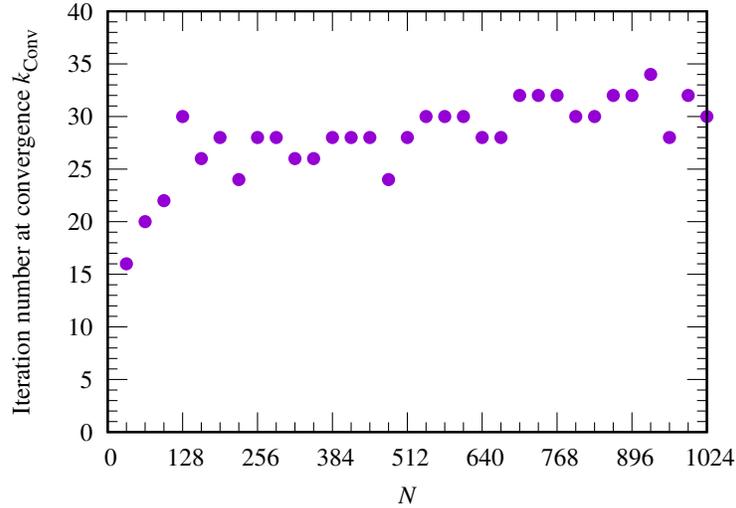}
    \caption{Relationship between the number of background grids $N$ and convergence iteration numbers $k_{\rm Conv}$.}
    \label{fig_compareN_IterNum}
  \end{center} 
\end{figure}

Fig. \ref{fig_compareTolerance_l_min} shows the relationships between relative displacement $e^{k}_{\infty}$ and the minimum distance between DPs $l_{\rm min}$ for different background meshes with respect to $N$.
The minimum distance $l_{\rm min}$ does not change remarkably for $e^{k}_{\infty} \leq 10^{-2}$ for each $N$, which infers that the DP positions converged well.
The convergence iteration numbers $k_{\rm Conv}$, at when $e^{k}_{\infty}$ reaches smaller than a convergence criterion $e_{\rm crit} = 10^{-2}$, are shown in Fig. \ref{fig_compareN_IterNum}.
When $N$ increases, $k_{\rm Conv}$ gradually increases for $N \geq 128$, but stays around $k_{\rm Conv} \approx 30$, which denotes practically acceptable convergence behavior.
Hereafter, the results are shown at $e_{\rm crit} = 10^{-2}$.

\begin{figure}[H] 
  \begin{center}
    \includegraphics[width=\linewidth]{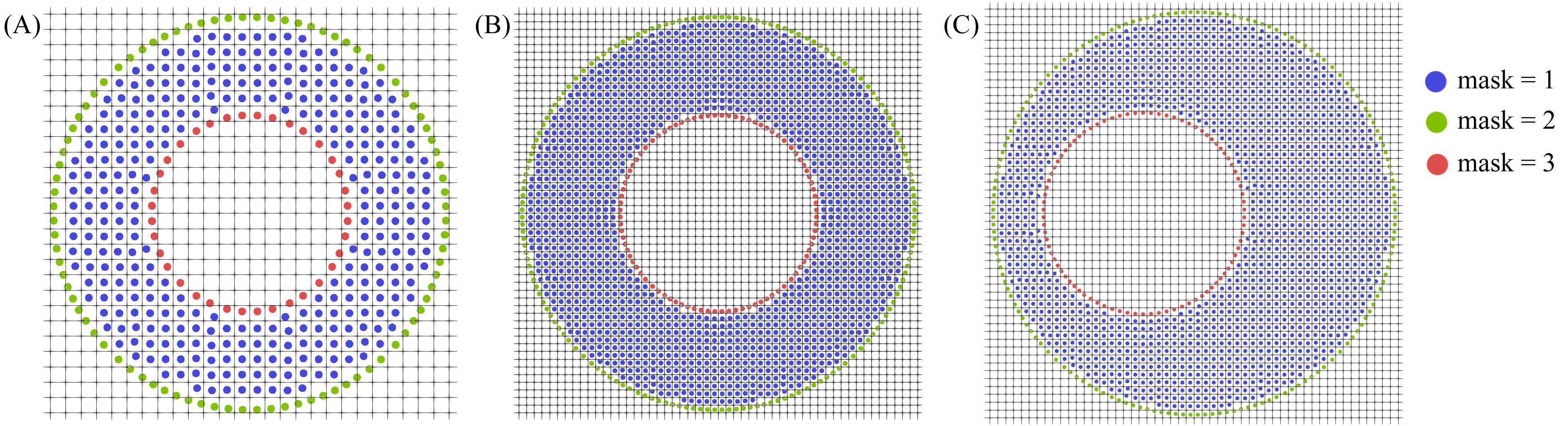}
    \caption{Distributions of DPs for co-axial circles at $N = 32$ (A) and $64$ (B) and the eccentric circle at $N = 64$ (C). The DPs are labeled as mask = 1 for the domain inside (blue dots), mask = 2 for the outer wall (green dots), and mask = 3 for the inner wall (red dots).}
    \label{fig_maskCCF_N32N64}
  \end{center}
\end{figure}

Fig. \ref{fig_maskCCF_N32N64} shows distributions of DPs in the cases for co-axial circles at $N = 32, 64$ and eccentric circles at $N = 64$, where the number of DPs for the domain inside $n_{\rm I}$, boundary $n_{\Gamma}$, and total $n_{\rm DP}$ are shown in Table \ref{table_numberDP}.
In all cases, the DPs demonstrate an even distribution, where the DPs with the boundary masks ($\geq 2$) lie on the boundary surfaces and the DPs with mask =1 are closely located at the mesh center.
Table \ref{table_lmin_lB} shows an apparent distance for the boundary DPs $l_{\Gamma} = 2\pi (r_{\rm O} + r_{\rm I}) / n_{\Gamma}$ and the minimum distance between DPs $l_{\rm min}$, where the relative differences from $l_{0}$ are also shown as $\varepsilon_{\Gamma} = (l_{\Gamma}-l_{0})/l_{0} \times 100$ and $\varepsilon_{\rm min} = (l_{\rm min} - l_{0})/l_{0} \times 100$, respectively.
The apparent distance for the boundary DPs $l_{\Gamma}$ is close to $l_0$ (or $h$), whereas the minimum distance $l_{\rm min}$ is smaller than $l_0$.
This indicates that the proposed MCD approach automatically provides an appropriate number of DPs on the boundary.

\begin{table}[H]
  \begin{center}
    \caption{Number of DPs for the domain inside $n_{\rm I}$, boundary $n_{\Gamma}$, and total $n_{\rm DP} (= n_{\rm I} + n_{\Gamma})$ in each case.}
    \begin{tabular}{lccc} \hline
      ~~ & $n_{\rm I}$ & $n_{\Gamma}$ & $n_{\rm DP}$ \\ \hline
      Co-axial circle ($N$ = 32) & 320 & 124 & 444 \\
      Co-axial circle ($N$ = 64) & 1420 & 248 & 1668 \\
      Eccentric circle ($N$ = 64) & 1418 & 246 & 1664 \\ \hline
    \end{tabular}
    \label{table_numberDP}
  \end{center}
\end{table}

\begin{table}[H]
  \begin{center}
    \caption{Comparisons of the apparent distances for the boundary DPs, $l_{\Gamma}$, and the minimum distance between DPs, $l_{\rm min}$, where the relative differences from $l_{0}$ are also given by $\varepsilon_{\Gamma} = (l_{\Gamma}-l_{0})/l_{0} \times 100$ and $\varepsilon_{\rm min} = (l_{\rm min} - l_{0})/l_{0} \times 100$.}
    \begin{tabular}{lcccc} \hline
      ~~ & $l_{\Gamma}$ & $\varepsilon_{\Gamma}$ (\%) & $l_{\rm min}$ & $\varepsilon_{\rm min}$ (\%) \\ \hline
      Co-axial circle ($N$ = 32) & $1.52 \times 10^{-1}$ & $-2.8$ & $1.31 \times 10^{-1}$ & $-16$ \\
      Co-axial circle ($N$ = 64) & $7.60 \times 10^{-2}$ & $-2.7$ & $6.57 \times 10^{-2}$ & $-16$ \\
      Eccentric circle ($N$ = 64) & $3.83 \times 10^{-2}$ & $-1.9$ & $3.18 \times 10^{-2}$ & $-19$ \\ \hline
    \end{tabular}
    \label{table_lmin_lB}
  \end{center}
\end{table}

\begin{figure}[H] 
  \begin{center}
    \includegraphics[width=90mm]{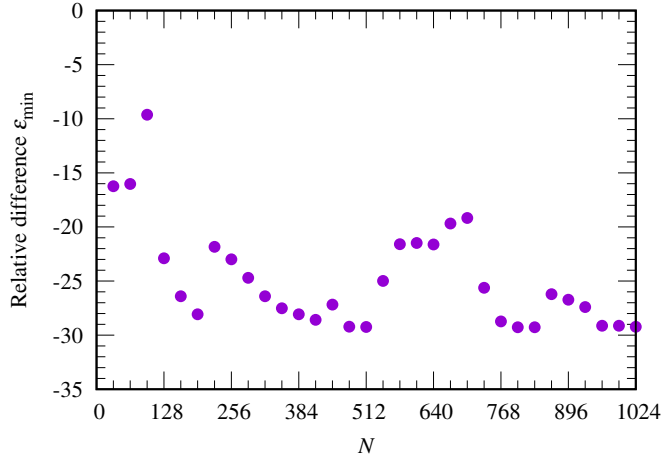} 
    \caption{Relationship between the number of background meshes $N$ and the relative difference of the minimum distance for DPs, $\varepsilon_{\rm min}$.}
    \label{fig_compareN_relErr} 
  \end{center} 
\end{figure}

The relationship between $\varepsilon_{\rm min}$ and $N$ is shown in Fig. \ref{fig_compareN_relErr}.
From Fig. \ref{fig_compareN_relErr}, it seems that $\varepsilon_{\rm min}$ tends to become negatively large as $N$ increases and converges to $\varepsilon_{\rm min} \approx -29\%$, which indicates that the minimum distance $l_{\rm min}$ becomes approximately $29\%$ smaller than the initial distance $l_0$ (or mesh width $h$). 

\begin{figure}[H] 
  \begin{center}
    \includegraphics[width=\linewidth]{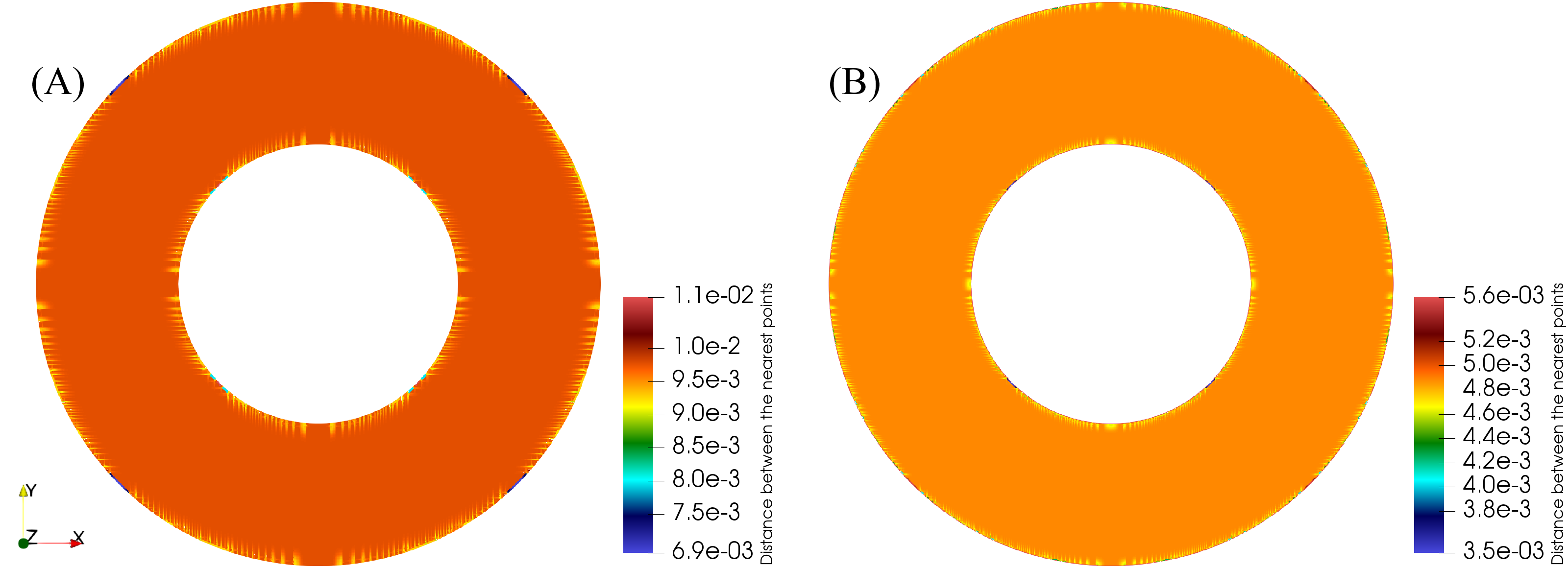}
    \caption{Distributions of $l_{\rm nearest}$ at $N = 512$ (A) and $1024$ (B), where each DP is colored using $l_{\rm nearest}$. Note that the initial distance $l_0$ is evaluated as $9.77 \times 10^{-3}$ for $N=512$ and $4.88 \times 10^{-3}$ for $N=1024$.}
  \label{fig_evalBoundary_lmin_N512_N1024}
  \end{center}
\end{figure}

\begin{figure}[H] 
  \begin{center}
    \includegraphics[width=\linewidth]{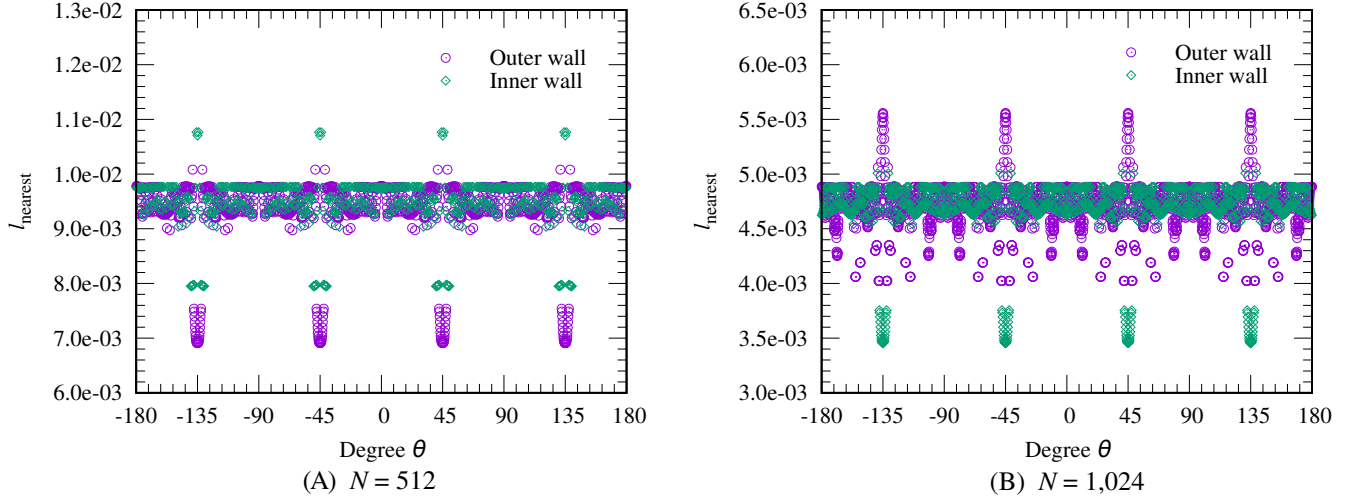} 
    \caption{Distributions of $l_{\rm nearest}$ on the inner and outer boundary walls at $N = 512$ (A) and $1024$ (B), where $\theta$ denotes the angle in degrees between the position vector of each DP and the $x$-axis. Note that $l_0/\sqrt{2} \approx 6.905 \times 10^{-3}$ for $N=512$ and $l_0/\sqrt{2} \approx 3.453 \times 10^{-3}$ for $N=1024$.}
    \label{fig_evalBoundary_N512_N1024}
  \end{center}
\end{figure}

\begin{figure}[H]
  \begin{center}
    \includegraphics[width=60mm]{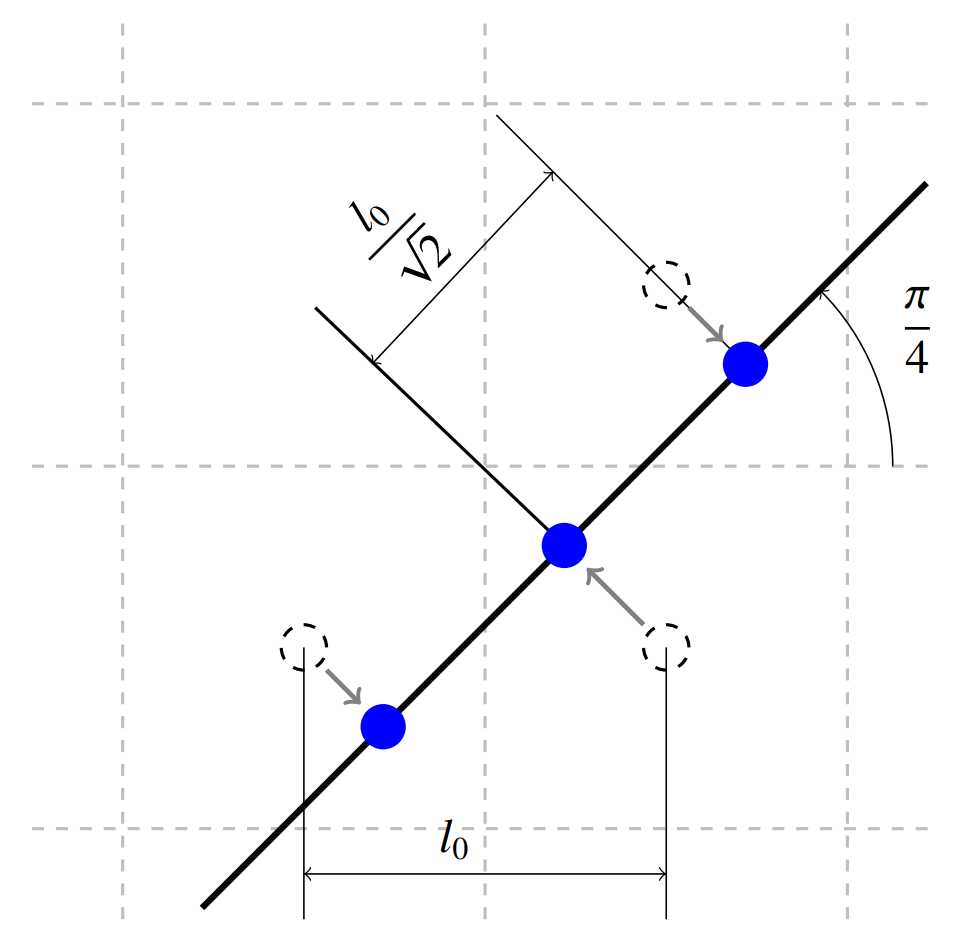} 
    \caption{Example in which $\varepsilon_{\rm min}$ asymptotically reaches $-29\%$ when $N \rightarrow \infty$.}
    \label{fig_reasonEpsilon-30}
  \end{center}
\end{figure}

To clarify where the minimum distance originates from and why it appears, the nearest distance for each DP, $l_{\rm nearest}$, is evaluated; that is, the local minimum distance to the surrounding DPs in $3 \times 3$ blocks for each DP.
Fig. \ref{fig_evalBoundary_lmin_N512_N1024} shows distributions of $l_{\rm nearest}$ at $N = 512$ and $1024$, where each DP is colored using $l_{\rm nearest}$.
In both meshes, the local distance $l_{\rm nearest}$ is nearly constant inside the domain and the value is similar to $l_0$, where $l_0=9.77 \times 10^{-3}$ for $N=512$ and $l_0=4.88 \times 10^{-3}$ for $N=1024$.
By contrast, variations exist around boundaries.
Fig. \ref{fig_evalBoundary_N512_N1024} shows distributions of $l_{\rm nearest}$ on the inner and outer boundary walls.
For both meshes, the local distance $l_{\rm nearest}$ reaches the minimum around the angles (from the $x$-axis) $\theta = \pm 135^{\circ}, \pm45^{\circ}$, and the values are close to $71\%$ of $l_0$. 
This can be understood using the following consideration. When the limit for $N \rightarrow \infty$ is taken, a boundary curve can be regarded as a straight line and the DPs have an arrangement such that the distances between DPs are $1/\sqrt{2} \approx 0.707$ times smaller than $l_{\rm 0}$, as shown in Fig. \ref{fig_reasonEpsilon-30}.
Thus, $\varepsilon_{\rm min}$ asymptotically approaches approximately $-29\%$ as $N$ increases, as shown in Fig. \ref{fig_compareN_relErr}.

It should be noted that the DPs distant from boundaries are closely located at mesh centers, which is attributed to short-range repulsion (\ref{eq_Fij_xij}), which only acts in the case when the DP distance is less than $l_0$.
This may achieve reasonable numerical accuracy for derivative evaluation because of the similarity to finite difference methods; however, the solutions may strongly depend on the background (Cartesian) meshes.
In this regard, in the current DP arrangement, the DP distance (or local spatial resolution) varies and becomes coarser near the boundaries.
Further consideration for an appropriate DP arrangement will be needed in the future.

\subsection{Circular Couette flow}
\label{circularCouetteFlow}

Fluid simulations are performed for a circular Couette flow problem using the distributions of DPs obtained in the previous section (Fig. \ref{fig_maskCCF_N32N64}).
The parameters are set to $\nu=0.1$, $\Delta t = 8 \times 10^{-4}$, $V_{\rm I}=0$, and $V_{\rm O}=1$, where $V_{\rm I}, V_{\rm O}$ denote the wall velocities on the inner and outer circular boundaries, respectively, taken as counterclockwise positive.
Note that the viscous stability indicator
\begin{equation}
  C_{\nu} = \frac{\nu \Delta t}{\left( l_{\rm min} \right)^{2}}
  \label{eq_nuStabilityCondition}
\end{equation}
is set to be sufficiently small in each spatial condition so that $C_{\nu} = 4.7 \times 10^{-3}$ ($N = 32$), $C_{\nu} = 1.9 \times 10^{-2}$ ($N = 64$), and $C_{\nu} = 8.8 \times 10^{-2}$ ($N = 128$).

In the derivative evaluation of velocity and pressure in Eqs. (\ref{eq_fStep1}), (\ref{eq_fStep2}), and (\ref{eq_fStep3}), the scaling parameter in the MLS reconstruction is set to $r_{\rm s} = l_{0}$.
Because the boundary treatment for incorporating the Neumann boundary condition (Matsunaga et al., 2020) is adopted, the velocity and pressure unknowns are solved for the DPs inside the domain, that is, $\textbf{x}_i \in \Omega_{\rm I}$, where the velocity on the boundary is directly given by the wall velocity as the Dirichlet boundary condition.
As previously described, $3 \times 3$ background grids are used for compact support $D_i$, and constant weight $w_{j}|_{i} = 1$.
The normal vector on the wall surface is similarly evaluated by the MLS reconstruction for the discrete SDFs, as described in Section \ref{makeSDF}, with the parameters $r^{\rm SDF}_{\rm e} = 3.1 h$ and $r^{\rm SDF}_{\rm s} = 0.7 h$.
The linear system for the pressure Poisson equation is solved using the Bi-CGSTAB method, assuming the system converges when the relative value of the residual vector norm to the initial value is below the tolerance set to $10^{-7}$.
The assumption is that the flow becomes a steady state when the instantaneous maximum velocity differences of the $x$ and $y$-components, $\max_{i}{ \left| \phi^{k}_{i} - \phi^{k-1}_{i} \right| }$ ($\phi=u,v$), are below $2 \times 10^{-8}$.

The theoretical (exact) solution is given by
\begin{equation}
	V_{\theta} = \frac{C_1}{r} + C_2 r,
	\label{eq_exactCCF}
\end{equation}
\begin{equation}
	C_1 = \left( \Omega_{\rm I} - \Omega_{\rm O} \right)
	\frac{r_{\rm I}^2 r_{\rm O}^2}{r_{\rm O}^2 - r_{\rm I}^2},~~
  C_2 = \frac{\Omega_{\rm O} r_{\rm O}^2 - \Omega_{\rm I} r_{\rm I}^2}
	{r_{\rm O}^2 - r_{\rm I}^2},
	\label{eq_C1,C2_exactCCF}
\end{equation}
where $V_{\theta}(r)$ is the tangential velocity, and $\Omega_{\rm O}=V_{\rm O}/r_{\rm O}$ and $\Omega_{\rm I}=V_{\rm I}/r_{\rm I}$ are the angular velocities of the outer and inner walls, respectively.
Because the pressure becomes constant and inherently undefined in this problem, $P^{\rm Exact} = 0$ is set.

\begin{figure}[H] 
  \begin{center}
    \includegraphics[width=0.9\linewidth]{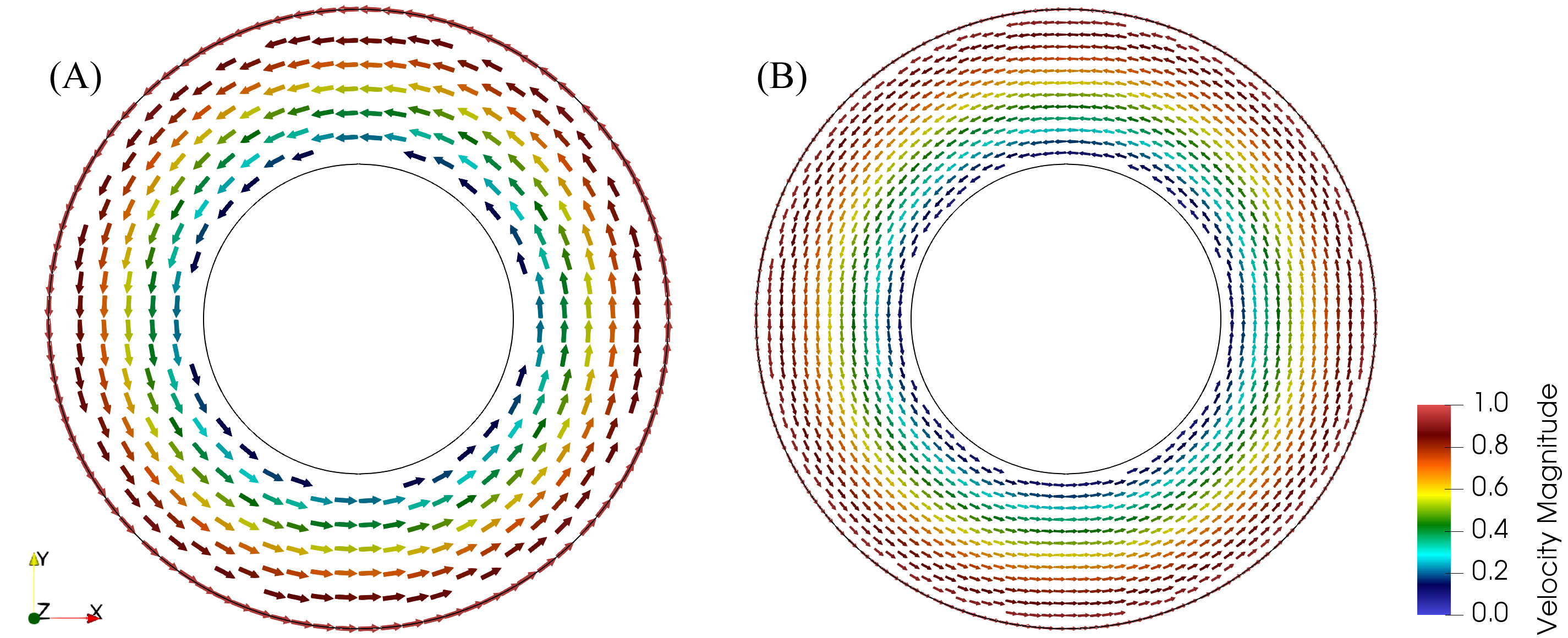}
    \caption{Numerical results for the velocity field in the circular Couette flow at $N = 32$ (A) and 64 (B).}
    \label{fig_vCCF_N32N64}
  \end{center}
\end{figure}

\begin{figure}[H] 
  \begin{center}
    \includegraphics[width=100mm]{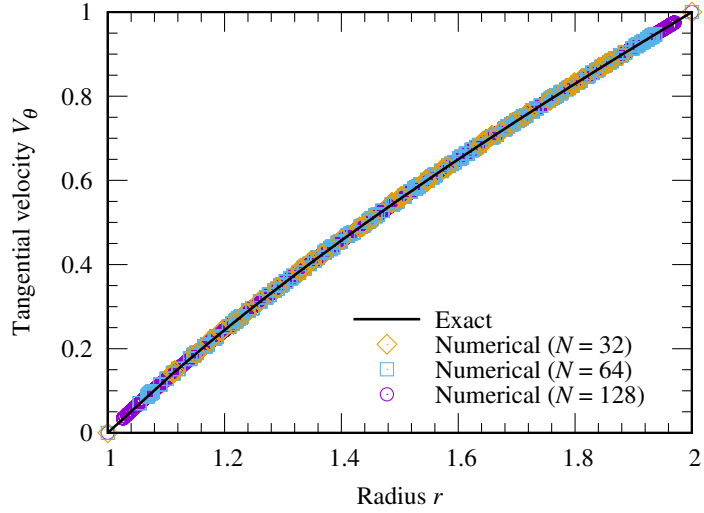} 
    \caption{Comparisons of the tangential velocity $V_{\theta}$ between numerical solutions at $N = 32, 64, 128$ and the exact solution along the radial direction $r$. The results are shown for all the DPs as $V_{\theta}={\bf v}\cdot{\bf e}_r$, where ${\bf e}_r$ is the unit basis vector in the radial direction $r$.}
    \label{fig_vAnaNumCCF_N32_N64}
  \end{center}
\end{figure}

The numerical results for the velocity field at $N=32$ and 64 are shown in Fig. \ref{fig_vCCF_N32N64} and comparisons for the tangential velocity $V_{\theta}$ between numerical solutions in $N = 32, 64, 128$ and the exact solution along the radial direction $r$ are shown in Fig. \ref{fig_vAnaNumCCF_N32_N64}.
The results are drawn at all DPs for each $N$.
Even for the coarser spatial resolution $N=32$, the results well reproduce a nonlinear profile of the velocity and are competitive with the exact profile.
Additionally, the effects of the mesh-based DP arrangements on the solutions seem to be negligible from the fact that circumferentially symmetric velocity profiles are well reproduced.

To check the numerical accuracy of the proposed method, $L_{1}$, $L_{2}$, and $L_{\infty}$ norms for numerical errors between the numerical solution $\phi$ and the exact solution $\phi^{\rm Exact}$ at the number of background meshes $N$ are evaluated as 
\begin{equation}
  L^{(N)}_{1}
  = \frac{1}{n_{\rm I}}
    \sum^{n_{\rm I}}_{i=1} 
    \left| \phi_{i} - \phi^{\rm Exact}_{i} \right|, 
\end{equation}

\begin{equation}
  L^{(N)}_{2} 
  = \sqrt{
    \frac{1}{n_{\rm I}}
    \sum^{n_{\rm I}}_{i=1}
    \left| \phi_{i} - \phi^{\rm Exact}_{i} \right|^{2}
  }
\end{equation}

\begin{equation}
  L^{(N)}_{\infty} 
  = \max_{ i \in [1,n_{\rm I}]}
  \left| \phi_{i} - \phi^{\rm Exact}_{i} \right|
\end{equation}
for $\phi=u,v,P$.
The numerical errors and convergence orders for $u, v, P$ are shown in Table \ref{table_SpatialConv}.
The order is calculated using $L^{(N)}_{p - \rm Order} = \log_{2} \left( L^{(N/2)}_{p} / L^{(N)}_{p} \right)$.
Note that the numerical pressure is shifted so that the average value in the domain inside $\Omega_{\rm I}$ becomes zero; that is, $\langle P \rangle = \sum^{n_{\rm I}}_{i=1} P_{i} / n_{\rm I} = 0$.
The convergence of the velocity achieves approximately second-order accuracy and the results between $u$ and $v$ are the same.
This demonstrates that a symmetric solution is obtained in the $x$ and $y$ directions for circular Couette flow.
For the pressure, although the accuracy of $L_{\infty}$ is slightly smaller than 2, particularly for the coarse spatial resolution, it almost achieves second-order accuracy.

\begin{table}[H]
  \begin{center}
    \caption{Numerical errors and convergence orders between the numerical result and exact result for $u$, $v$, and $P$.}
    \begin{tabular}{llcccccc} \hline 
    ~~ & $N$ & $L_{1}$ & ~~$L_{1-\rm Order}$~~ & $L_{2}$ & ~~$L_{2-\rm Order}$~~ & $L_{\infty}$ & ~~$L_{\infty - \rm Order}$~~  \\ \hline
    $u$~~ & 32      & ~~$1.31 \times 10^{-3}$~~ & --   & ~~$1.73 \times 10^{-3}$~~ & --   & ~~$4.85 \times 10^{-3}$~~ & --   \\
    ~~ & 64      & ~~$3.10 \times 10^{-4}$~~ & 2.08 & ~~$4.08 \times 10^{-4}$~~ & 2.08 & ~~$1.07 \times 10^{-3}$~~ & 2.18 \\
    ~~ & 128~~ & ~~$8.18 \times 10^{-5}$~~ & 1.92 & ~~$1.05 \times 10^{-4}$~~ & 1.96 & ~~$2.70 \times 10^{-4}$~~ & 1.99 \\
    & & & & & & & \\
    $v$~~ & 32      & ~~$1.31 \times 10^{-3}$~~ & --   & ~~$1.73 \times 10^{-3}$~~ & --   & ~~$4.85 \times 10^{-3}$~~ & --   \\
    ~~ & 64      & ~~$3.10 \times 10^{-4}$~~ & 2.08 & ~~$4.08 \times 10^{-4}$~~ & 2.08 & ~~$1.07 \times 10^{-3}$~~ & 2.18 \\
    ~~ & 128~~ & ~~$8.18 \times 10^{-5}$~~ & 1.92 & ~~$1.05 \times 10^{-4}$~~ & 1.96 & ~~$2.70 \times 10^{-4}$~~ & 1.99 \\
    & & & & & & & \\
    $P$~~ & 32      & ~~$2.83 \times 10^{-3}$~~ & --   & ~~$3.80 \times 10^{-3}$~~ & --   & ~~$1.39 \times 10^{-2}$~~ & --   \\
    ~~ & 64      & ~~$4.41 \times 10^{-4}$~~ & 2.68 & ~~$8.20 \times 10^{-4}$~~ & 2.21 & ~~$6.19 \times 10^{-3}$~~ & 1.17 \\
    ~~ & 128~~ & ~~$9.68 \times 10^{-5}$~~ & 2.19 & ~~$1.62 \times 10^{-4}$~~ & 2.34 & ~~$1.85 \times 10^{-3}$~~ & 1.74 \\ \hline
    \end{tabular}
    \label{table_SpatialConv}
  \end{center}
\end{table}

\subsection{Eccentric circular Couette flow}
\label{eccentricCircularCouetteFlow}

As further validation, the eccentric circular Couette flow problem is solved using the DPs obtained in Section \ref{distributionDP}. The geometrical configuration and distribution of DPs at $N=64$ are shown in Table \ref{table_dimension} and Fig. \ref{fig_maskCCF_N32N64} (C).
The parameters are set to $\nu=0.1$, $\Delta t = 1.5 \times 10^{-3}$, $V_{\rm I}=0$, and $V_{\rm O}=-1$.
The viscous stability indicator (\ref{eq_nuStabilityCondition}) becomes $C_{\nu} = 7.6 \times 10^{-2}$.
The numerical setup for the MLS reconstruction and pressure Poisson equation are the same as those for circular Couette flow, except the convergence criterion for the velocity differences used to identify the steady state is set to $10^{-6}$.

\begin{figure}[H] 
  \begin{center}
    \includegraphics[width=120mm]{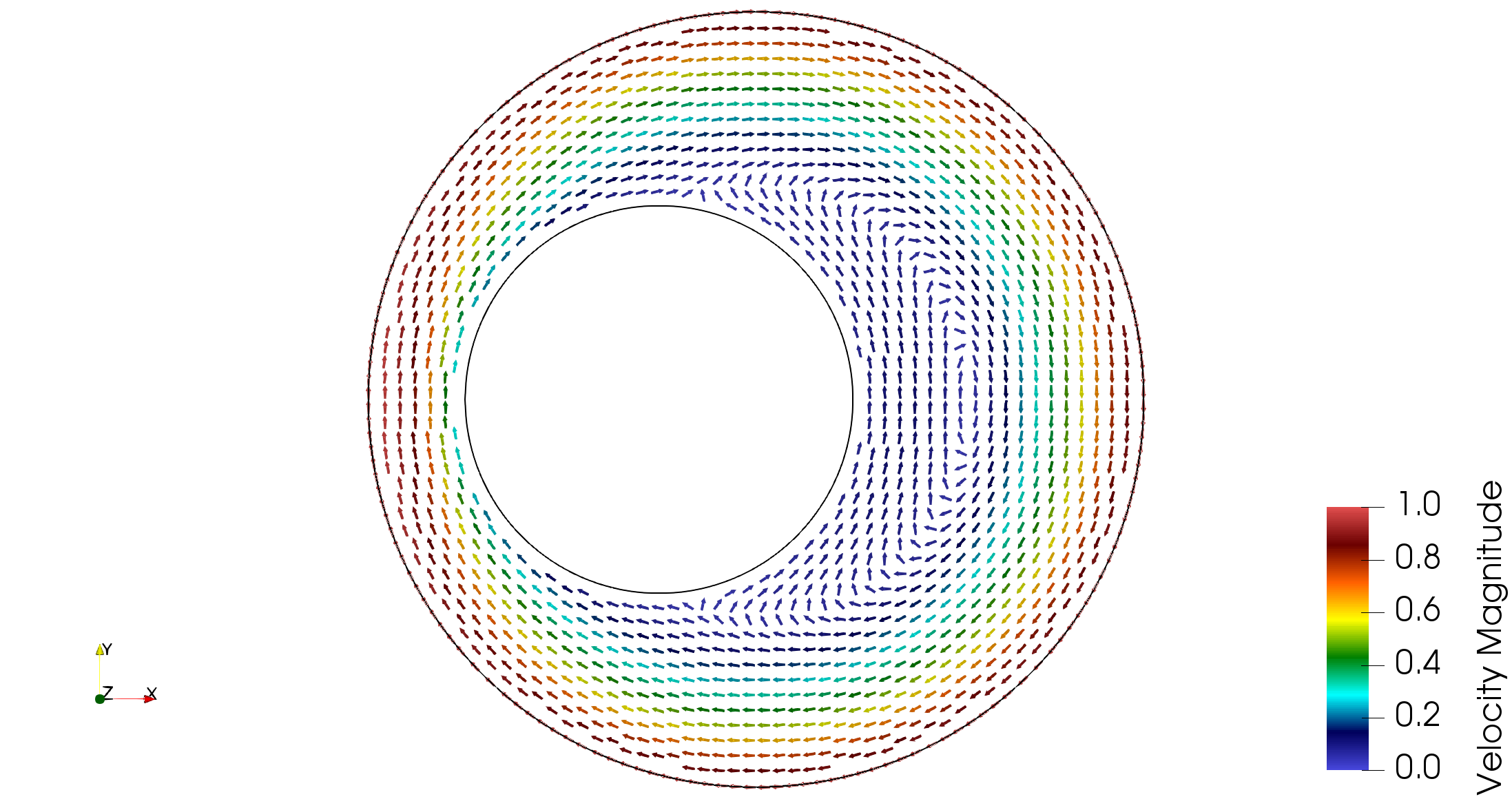}
    \caption{Numerical velocity field for eccentric circular Couette flow.}
  \label{fig_vECCF_N64}
  \end{center}
\end{figure}

Fig. \ref{fig_vECCF_N64} shows the numerical solution for the velocity field.
Because of the eccentric circular arrangement, the velocity field does not demonstrate cylindrically symmetric flow and the flows swirl in the opposite direction at the right-hand side of the inner wall with respect to the outer flows.
This flow pattern is well known for flows in the eccentric circular channel.

\begin{figure}[H] 
  \begin{center}
    \includegraphics[width=\linewidth]{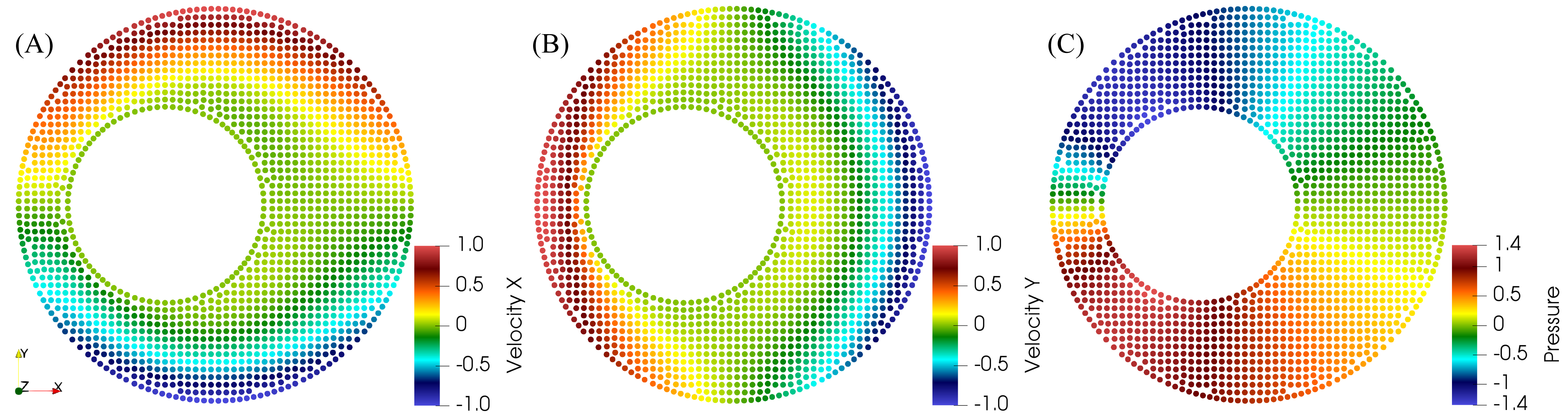}
    \caption{Numerical results for velocity components $u$ (A), $v$ (B), and pressure $P$ (C) in eccentric circular Couette flow. The boundary pressure is reproduced from the obtained discrete pressures inside the domain using the MLS reconstruction described in Appendix \ref{appendixA}, where the parameters are set to $r^{bd}_{\rm s} = l_{0}$ and $r^{bd}_{\rm e} = 3.1 l_{0}$ with the weight function (\ref{eq_weight}).}
  \label{fig_vxyECCF_N64}
  \end{center}
\end{figure}

\begin{figure}[H] 
  \begin{center}
    \includegraphics[width=100mm]{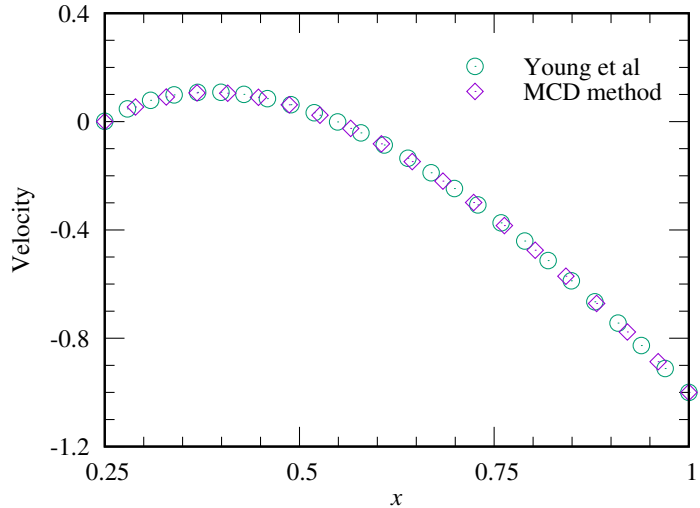} 
    \caption{Comparison of velocity profiles between the present MCD method and the dual-potential formulation (Young et al., 2006). The velocity distribution shows the axial profile of $u$ at $y = 0$, where the reference solution is extracted from the literature and the present result is interpolated using the MLS reconstruction presented in this paper.}
    \label{fig_ComparisonYoung_N64}
  \end{center}
\end{figure}

Fig. \ref{fig_vxyECCF_N64} shows the distributions of velocity components $u,v$ and pressure $P$.
All the quantities are smoothly obtained at each DP, and the velocity components are in good agreement with those solved using a well-validated numerical method (Zhang and Zhang, 2014).
For the sake of visualization, the boundary pressure is extrapolated from the obtained discrete pressures inside the domain using the MLS reconstruction described in Appendix \ref{appendixA}.
Note that this extrapolation does not affect numerical results because the boundary pressures are not required in this analysis.
In Fig. \ref{fig_ComparisonYoung_N64}, the axial profile of $u$ at $y=0$ for $0.25 \leq x \leq 1$ is compared with that provided by the dual-potential formulation (Young et al., 2006), where the present results are interpolated by the MLS reconstruction presented in this paper.
The result excellently captures the reference solution.

\section{Conclusions}
In this study, a novel approach was proposed for the use of compact stencils in particle-based meshless methods, called the MCD approach.
The DPs were linked to the background Cartesian meshes and the positions were moved by solving the dynamic problem with nonlinear kinematic conditions. 
As a result, each DP was rigorously constrained in each background mesh, and the MLS reconstruction was performed using the local DPs for $3 \times 3$ background grids in two dimensions.

Numerical tests were conducted for the co-axial and eccentric circular flows governed by the two-dimensional Stokes equations.
The results confirmed that the proposed algorithm for DP arrangements was stable for $\kappa \leq 1/3$ and provided an entirely homogeneous distribution of DPs that was independent of spatial resolutions, where the minimum DP distance near the boundary asymptotically approached $71\%$ of $l_{0}$ when the spatial resolutions increased.
From the flow analyses using the obtained DP arrangements, the present MCD obtained accurate solutions for both flow problems, where the results achieved approximately second-order accuracy for the velocity and pressure, as expected, and were competitive with those of existing methods.

Although particle-based meshless methods underlie the present method, the proposed MCD approach achieved compact stencils with a regular distribution ($3 \times 3$ in two dimensions).
This could bring the following advantages: (1) Easy application for the moderate flows with arbitrary boundary shapes without mesh generation procedures; (2) Low computational cost, which would be competitive to that of traditional Cartesian mesh systems, because of the compactness and equality of stencils in derivative evaluations on each DP.
This would bring another advantage of high parallel efficiency in large-scale simulations.
Because the DPs are linked to the background meshes, any acceleration solver for the linear system can be applied, such as a bucket-based multigrid preconditioner used in a conventional particle method (S\"{o}dersten et al., 2019).
Moreover, the proposed MCD method could be coupled with a highly efficient parallelization technique using the Cartesian grid system (Jansson et al., 2019).

In addition, the proposed method has potential advantages in moving boundary problems. In general, material points near the largely moved and deformed interface are highly changed over time, and thus the neighboring relation on each particle is dynamically changed during calculation. The proposed mesh-constrained idea is possible to only focus on the background meshes, which manage the DPs, and would avoid suffering from the local neighboring changes of DPs.

Conclusively, the present MCD method is an accurate and practical approach, even in a particle-based meshless method, and can be used for solving two-dimensional Stokes flows.
Regarding future studies, extensions to the Navier-Stokes equations, a more sophisticated DP arrangement, higher-order accurate formulation, and moving boundary problems with DP movement will be considered. Also, the current formulation does not guarantee the conservation properties in discrete level, and hence the improvement will be required to apply for the practical problems.

\section*{Acknowledgements}
This research was supported by JSPS KAKENHI grant No. JP19H01175 and JP22K19939; MEXT as a ``Program for Promoting Researches on the Supercomputer Fugaku'' (hp210181, hp220161); the High-Performance Computing Infrastructure System Research Project (hp210033, hp220106); and Tokyo Metropolitan Government (Grant No. R2-2).

\appendix

\begin{appendices}
\section*{Appendix}

\section{MLS reconstruction on the Neumann boundary $\Gamma_{\rm N}$}
\label{appendixA}
An interpolation is explained for an arbitrary variable at $\textbf{x}_c \in \Gamma_{\rm N}$.
The objective function is extended from Eq. (\ref{eq_objFunc}) as follows:
\begin{equation}
  J = \frac{1}{2} \!\!\! \sum_{
      {\fontsize{7pt}{0mm}\selectfont
        \begin{array}{l}
          j \! \in \! \Lambda_c
        \end{array}}
      }{\!\!\!
      w_j \left( \textbf{p}_j \cdot \tilde{\bm{\Phi}} +\phi_c -\phi_j \right)^2}
    + \frac{1}{2} \!\!\! \sum_{
      {\fontsize{7pt}{0mm}\selectfont
        \begin{array}{l}
          j \! \in \! \Lambda^{\rm N}_c
        \end{array}}
      }{\!\!\!
      w_j \left\{ r_{\rm s} \left( \frac{1}{r_{\rm s}} \textbf{p}^{\rm N}_j \cdot \tilde{\bm{\Phi}} -f_{j} \right) \right\}^{2}}
    + \lambda^{\rm N} \left( \textbf{p}^{\rm N}_c \cdot \tilde{\bm{\Phi}} - r_{\rm s} f_{c} \right),
  \label{ap_eq_objFunc}
\end{equation}
where $\textbf{p}^{\rm N}_c = \textbf{p}^{\rm N} \left( \textbf{X}_c = 0 \right)$ and $f_{c} = f(\textbf{x}_c)$.
Note that the last term in Eq. (\ref{eq_objFunc}) is ignored in this study.
The third term represents the constraint for the Neumann boundary condition with the introduction of the Lagrange multiplier $\lambda^{\rm N}$.

The stationary conditions are derived with respect to $\phi_c$, $\tilde{\bm{\Phi}}$ and $\lambda^{\rm N}$:

\begin{equation}
  \begin{array}{l}
    \displaystyle ~~~~~
      \frac{\partial J}{\partial \phi_c} = 0,~~
      \frac{\partial J}{\partial \tilde{\bm{\Phi}}} = \bm{0},~~
      \frac{\partial J}{\partial \lambda^{\rm N}} = 0, 
    \quad \Rightarrow \quad
    \left\{
      \begin{array}{l}
        \displaystyle 
          a \phi_c + \textbf{b} \cdot \tilde{\bm{\Phi}}
          = c, \\
        \displaystyle 
          \textbf{b} \phi_c + \left(\textbf{L}+\textbf{L}^{\rm N}\right) \tilde{\bm{\Phi}}
          + \lambda^{\rm N} \textbf{p}^{\rm N}_{c}
          = \textbf{d} + \textbf{d}^{\rm N}, \\
        \displaystyle 
          \textbf{p}^{\rm N}_{c} \cdot \tilde{\bm{\Phi}} 
          = r_{\rm s} f_{c},
      \end{array}
    \right.
  \end{array}
    \label{ap_eq_simulations}
\end{equation}
where $a$, $\textbf{b}$, $c$, $\textbf{d}$, $\textbf{d}^{\rm N}$, $\textbf{L}$,  and $\textbf{L}^{\rm N}$ are given in Eqs. (\ref{eq_a})--(\ref{eq_L}), respectively.
By eliminating $\phi_c$, Eq. (\ref{ap_eq_simulations}) can be written as

\begin{equation}
  \left\{
    \begin{array}{l}
      \displaystyle 
      \textbf{M} \tilde{\bm{\Phi}} 
      + \lambda^{\rm N} \textbf{p}^{\rm N}_{c}
        = \textbf{e}, \\
      \displaystyle 
        \textbf{p}^{\rm N}_{c} \cdot \tilde{\bm{\Phi}} 
        = r_{\rm s} f_{c},
    \end{array}
  \right.
  \label{ap_eq_simulations2}
\end{equation}
where $\textbf{M}$ and $\textbf{e}$ are defined in Eqs. (\ref{eq_M2}) and (\ref{eq_e2}).
By further eliminating $\tilde{\bm{\Phi}}$, the following equation is derived for $\lambda^{\rm N}$:

\begin{equation}
  \lambda^{\rm N}
  = \frac{
    \textbf{p}^{\rm N}_{c} \cdot \textbf{h} - r_{\rm s} f_{c}
    }{
      \textbf{p}^{\rm N}_{c} \cdot \textbf{h}^{\rm N}
     },
  \label{ap_eq_lmd}
\end{equation}
where
\begin{equation}
  \textbf{h}
  = \textbf{M}^{-1} \textbf{e}, \ \
  \textbf{h}^{\rm N} 
  = \textbf{M}^{-1} \textbf{p}^{\rm N}_{c}.
\end{equation}
Then, the modal components $\tilde{\bm{\Phi}}$ can be obtained as $\tilde{\bm{\Phi}} = - \lambda^{\rm N} \textbf{h}^{\rm N} + \textbf{h}$.

\end{appendices}

\end{document}